\documentclass[11pt]{amsart}
\usepackage{geometry}                
\usepackage[parfill]{parskip}    
\usepackage{graphicx}
\usepackage{amssymb}
\usepackage{epstopdf}

\theoremstyle{plain}
\newtheorem{thm}{Theorem}[section]
\newtheorem{theorem}{Theorem}[section]

\theoremstyle{definition}
\newtheorem{cor}[thm]{Corollary}

 \newtheorem{remark}{Remark}
 \newtheorem{lemma}[theorem]{Lemma}

 \newtheorem{prop}[thm]{Proposition}

\title{Caustics, Counting Maps and Semi-Classical Asymptotics}
\author{N. M. Ercolani}%
\address{Dept. of Math, Univ. of Arizona, 520-621-2713, FAX: 520-626-5186}%
\email{ercolani@math.arizona.edu}%
\thanks{The support of the Natonal Science Foundation, DMS-0808059, is gratefully acknowledged.}
\begin{document}

\begin{abstract}
This paper develops a deeper understanding of the structure and combinatorial significance of the partition function for Hermitian random matrices. The coefficients of the large $N$ expansion of the logarithm of this partition function, also known as the {\it genus expansion},  (and its derivatives) are generating functions for a variety of graphical enumeration problems. The main results are to prove that these generating functions are in fact specific rational functions of a distinguished irrational (algebraic) function, $z_0(t)$, of the generating function parameter, $t$. This distinguished function is itself the generating function for the Catalan numbers (or generalized Catalan numbers, depending on the choice of weight of the parameter). It is also a solution of the inviscid Burgers equation for certain initial data. The shock formation, or {\it caustic}, of the Burgers characteristic solution is directly related to the poles of the rational forms of the generating functions. 

As an intriguing application, one gains new insights into the relation between certain derivatives of the genus expansion, in a {\it double-scaling limit}, and the asymptotic expansion of the first Painlev{\'e} transcendent. This provides a precise expression of the Painlev{\'e} asymptotic coefficients directly in terms of the coefficients of the partial fractions expansion of the rational form of the generating functions established in this paper.  Moreover, these insights point toward a more general program relating the first Painlev{\'e} hierarchy to the higher order structure of the double-scaling limit through the specific rational structure of generating functions in the genus expansion.  The paper closes with a discussion of the relation of this work to recent developments in understanding the asymptotics of graphical enumeration. 

As a byproduct, these results also yield new information about the asymptotics of recurrence coefficients for orthogonal polynomials with respect to exponential weights, the calculation of correlation functions for certain tied random walks on a 1-D lattice, and the {\it large time} asymptotics of random matrix partition functions. 
\end{abstract}

\maketitle

\section{Introduction} \label{intro} This paper describes the coefficients in an asymptotic expansion of the logarithm of the partition function, $Z_N(t)$, 
$$
 \log \left(Z_{N}(t)/(Z_{N}(0)\right) = N^{2} e_{0}({t}) + e_{1}({{t}}) + \frac{1}{N^{2}} e_{2}(t) + \cdots +  \frac{1}{N^{2g-2}} e_{g}(t) + \cdots
$$
for certain ensembles of random Hermitian matrices. (See Theorem \ref{EQMSTHM} for a precise statement.)  

As is familiar from general contexts of statistics and probability theory,  these coefficients contain a great deal of information about expectations and correlation functions with respect to the Gibbs measure that underlies the given random matrix ensemble.  In our context we are interested in the asymptotic behavior of $N^{-2}\log Z_{N}(t)$ (often called the {\it free energy}) as the size, $N$, of the matrices becomes large.  These ensembles, and hence the asymptotic coefficients $e_g$, as well, depend on a coupling parameter $t$ (the analogue of the inverse temperature in the classical setting of the canonical ensemble). It is natural to view the $e_g(t)$ as generating functions for the just-mentioned expectations and correlation functions and therefore it is of value to know as much as possible about how these coefficients depend on the coupling parameters. 

We will derive general closed form expressions for {\it all} of the coefficients, viewed as generating functions, in our asymptotic expansions as well as in certain derivatives of these expansions.  We will show that, with only a couple exceptions, they are all rational functions of a single irrational (algebraic) function of $t$ which we denote by $z_0$. More succinctly, we will see that, for $g \geq 2$,
$$
e_g(t) \in \mathbb{Q}\left(z_0(-t)\right),\,\, \mbox{the function field of}\,\, z_0(-t) \,\, \mbox{over the rational numbers}\,\, \mathbb{Q}.   
$$ 
It is a standard and fairly elementary fact that a generating function is rational if and only if its sequence of coefficients eventually (i.e., beyond a certain order) satisfies a {\it linear} recurrence relation \cite{Lando}. The generating functions, such as $e_g(t)$, that we will describe are not rational in $t$; however, the result just mentioned suggests that if $e_g$ is indeed rational in $z_0$, then it should in fact be possible to recursively construct it from $z_0$.  This is, in effect, the viewpoint which guided our derivation of the rational form of the $e_g$ and other generating functions that arise as coefficients in asymptotic expansions  closely related to our partition function expansions (see Theorems \ref{egratl} and \ref{result}). For the fundamental function, $z_0(-t)$ itself, we are able to describe it very explicitly from several different perspectives. For instance, it is itself the generating function for {\it generalized Catalan numbers} which are familiar constructs in {\it enumerative combinatorics}.
\medskip

Putting all this together, our results provide a fairly detailed and systematic understanding of the asymptotic behavior of correlation functions for the random matrix ensembles that we study.  While this is interesting in and of itself, there is another level of applications for these results. The correlation functions that we are discussing also have a combinatorial interpretation related to {\it diagrammatic expansions} that are familiar in statistical physics. 

The $e_g(t)$ can be topologically identified as generating functions  that count labelled maps.  Roughly speaking, a  {\it map} is an embedding of an oriented graph into a compact, oriented and connected  surface $X$ with the requirement that the complement of the graph in the surface should be a disjoint union of simply connected open sets. A labelled map is a map on which the vertices of the embedded graph have been ordered. (Two such labelled maps are equivalent if the respective graphs can be smoothly deformed into one another within the surface.) The $n^{th}$ Taylor coefficient of $e_g(t)$ is an integer equal to the number of distinct equivalence classes of maps to a compact, oriented Riemann surface of genus $g$ that can be constructed using exactly $n$ vertices and where each vertex has the same valence. (This valence is given by a weight associated to the coupling parameter. See Section \ref{diffweights} and Theorem \ref{thm:01}.) The results presented in this paper yield recursion formulae for the coefficients of these generating functions, now viewed as counting functions for labelled maps, that are  based on the partial fractions expansions of the rational forms of the $e_g$ and related generating functions. 

Of particular interest is the new light that this application sheds on a 20 year old problem stemming from the physics literature on 2D Quantum Gravity \cite{BreKaz, DoS, GrMi}. In mathematical terms this problem asks for the detailed description of a particular {\it double-scaling limit} of the random matrix partition function.  We refer the reader to \cite{FIKN} for the history and background. We can give an idea of the contribution of our work to this problem by mentioning that $z_0(-t)$ is the profile of the self-similar solution to an inviscid Burgers equation (see Section \ref{IBE}). From the method of characteristics one may conclude that $z_0(-t)$ has an analytic continuation  to the complex $t$-plane minus a branch cut that extends from 
$-\infty$ to a critical value $-t_c < 0$ along the negative real axis. This critical time is in fact the {\it shock time} at which the Burgers caustic starts to form.  One then defines the double-scaling limit to be the  limit in which $t \to -t_c$ within a sector of the complex plane with $\Re t >  -t_c$ while at the same time 
$N \to \infty$ so that the similarity variable 
\begin{eqnarray*}
 \xi &=& \frac{N^{4/5} (t + t_c)}{-\gamma},
\end{eqnarray*} 
remains  fixed at some finite, non-zero value ($\gamma$ is a constant to be determined).  
Because of our rationality result, we are, first of all, able to conclude that the coefficients, $e_g(t)$, all have the same envelope of holomorphy as $z_0(-t)$, which is the above mentioned slit plane with a common singularity at $-t_c$. But, more strikingly, what we find in the double scaling limit, is that the dependence on the combinatorial information in $z_0(-t)$ washes out, so that all that remains in the limit is the information that can be deduced from the structure of $e_g$ as a rational function of $z_0$,  and that rational structure is what we describe in this paper.  

One may study the asymptotics of the free energy directly in the double-scaling limit.  This in fact was done, for a particular ensemble, in \cite{FIKI, FIKII} (see also \cite{DK})  via a Riemann-Hilbert representation of polynomials orthogonal with respect to the exponential weight that determines the Gibbs measure for the matrix ensemble under consideration. They find that this Riemann-Hilbert problem can be deformed into the Riemann-Hilbert problem for the first Painlev{\'e} transcendent (PI) with respect to the parameter $\xi$ in the double-scaling limit. Through this they are, roughly speaking, able to relate the double scaling limit of  $N^{-2}\log Z_N$ to an asymptotic expansion of PI in a sector of the $\xi$ plane. 

Our approach is also based on the Riemann-Hilbert problem for orthogonal polynomials, but  we develop explicit expressions for the double scaling asymptotics of the partition function, for a very general class of ensembles, in terms of our representations of the map counting functions as rational functions of $z_0$. This is consistent with the results for the special case considered in \cite{FIKI, FIKII} and therefore gives, for the first time as far as we aware, a {\it direct and explicit} interpretation of the coefficients for the PI asymptotic expansion in terms of the graphical enumeration of labelled maps. This connection opens up some interesting new directions for investigation that will be briefly discussed in section \ref{dsII}.

The rest of this introduction will amplify upon and detail the themes that have just been sketched.

\subsection{Partition Functions for Unitary Ensembles} \label{PFUE} The ensembles of interest in this paper are a sub-family of what are known as the Unitary Ensembles (UE) of random matrices; the common space of these ensembles is the space, $\mathcal{H}_n$, of $n \times n$ Hermitian matrices, $M$. The Gibbs measures for the ensembles we consider are given by the following family of probability measures,  
\begin{eqnarray} \label{RMT}
d\mu_{t_{2\nu}} &=& \frac{1}{\widetilde{Z}^{(n)}_N(t_{2\nu})}\exp\left\{-N \mbox{ Tr} [V_\nu(M, t_{2\nu})]\right\} dM,\,\, \mbox{where}\\
\label{I.001b} V_\nu(\lambda; \ t_{2\nu} ) &=&  \frac{1}{2} \lambda^{2} +  t_{2\nu} \lambda^{2\nu}
\end{eqnarray}
and the parameter $t_{2\nu}$ is assumed to be such that the integral converges.  For example, one may suppose that $\Re t_{2\nu} > 0$. 

The normalization factor  $\widetilde{Z}^{(n)}_N(t_{2\nu})$, which serves to make $\mu_t$ be a probability measure, is called the {\it partition function} of this unitary ensemble.  Because the trace, Tr, is invariant under conjugation and since $M$ is Hermitian, this partition function reduces to an expression which is proportional to the following multivariate density expressed in terms of the eigenvalues $\left\{\lambda_j\right\}$ of  $M$  \cite{Mehta}.

\begin{eqnarray}
\label{I.001}
& & Z^{(n)}_{N}(t_{2\nu}) =\\
\nonumber & & \int \cdots \int \exp{ \left\{
-N^{2}\left[\frac{1}{N} \sum_{j=1}^{n} V_\nu(\lambda_{j}; \ t_{2\nu})  - \frac{1}{N^{2}} \sum_{j\neq \ell} \log{|
\lambda_{j} -
\lambda_{\ell} | } \right] \right\} } d^{n} \lambda.
\end{eqnarray}
${Z}^{(n)}_N(t)$ differs from $\widetilde{Z}^{(n)}_N(t)$ only by an overall factor which is independent of $t$. (That is because the reduction to eigenvalues comes from a change of variables on $M$ that conjugates it to a diagonal matrix; given the invariance of the trace under conjugation, the effect of this variable change is only seen in $dM$ and does not involve $t$.) In all our considerations, we will always be working with ${Z}^{(n)}_N(t)/{Z}^{(n)}_N(0)$. So from this perspective, ${Z}^{(n)}_N(t)$ and $\widetilde{Z}^{(n)}_N(t)$ are equivalent. 

For the coupling weight $V_\nu$, we will henceforth often drop the subscript $\nu$ and just write $V$; similarly, for the coupling parameter $t_{2\nu}$ we will usually drop the subscript $2\nu$ and just write $t$. Which value of $\nu$ is intended should be clear from context. (However see section \ref{diffweights} for further comment.)
\medskip

In studying asymptotic questions we will always assume that $N$ and $n$ go to infinity together such their ratio $x = n/N$ remains constant at a finite, non-zero value.  When $t=0$, the Gibbs measure (\ref{RMT}) is the measure for the Gaussian Unitary Ensemble (GUE). We will typically want to work with a rescaling of our matirx ensembles given by $A = \sqrt{N}M$ and which we will refer to as the {\it fine scale}. In terms of this calibration, the probability measure at $t=0$ is

\begin{eqnarray} \label{GUE}
d\mu_\textbf{0}(A) &=& 2^{-n/2} \pi^{-n^2/2}\exp\left\{- 1/2 \mbox{ Tr} [A^2]\right\} dA.
\end{eqnarray} 
The general partition function associated to the measure $\mu_t$, or equivalently the integral (\ref{I.001}), can be expressed as an expectation with respect to the GUE ensemble. First observe that

\begin{eqnarray} \label{GUEEXP}
 \widetilde{Z}^{(n)}_{N}(t) &=& \int_{\mathcal{H}_n}  \exp\left\{- N \,\, \mbox{Tr} [1/2 \, M^2 + t M^{2\nu}]\right\} dM\\
\nonumber &=& N^{-n^2/2} \int_{\mathcal{H}_n}  \exp\left\{-  \,\, \mbox{Tr} [1/2 \, A^2 + \frac{t}{N^{\nu-1}} A^{2\nu}]\right\} dA\\
\nonumber &=& 2^{n/2} \left(\frac{\pi}{N}\right)^{n^2/2}  \,\, \mathbb{E}_{\mu_0}\left( \exp\left\{- \frac{t}{N^{\nu-1}}\mbox{Tr} [A^{2\nu}]\right\}\right)
\end{eqnarray}
where $\mathbb{E}_{\mu_0}$ denotes the expectation with respect to the measure (\ref{GUE}). It is then natural to define a function that interpolates naturally between the matrix and eigenvalue representations of the partition functions:

 \begin{eqnarray} \label{GUEEXPNorm}
\tau^2_{n,N}(t) =  \frac{Z^{(n)}_{N}(t)}{Z^{(n)}_{N}(0)} &=& \frac{\widetilde{Z}^{(n)}_{N}(t)}{\widetilde{Z}^{(n)}_{N}(0)}
 = \mathbb{E}_{\mu_0}\left( \exp\left\{- \frac{t}{N^{\nu-1}}\mbox{Tr} [A^{2\nu}]\right\}\right),
\end{eqnarray} 
\bigskip
where $\tau_{n,N}$ is called the $n^{th}$ scaled {\it tau function}. The {\it unscaled} tau function is $\tau_{n,1}$.

A primary motivation for carrying out the detailed asymptotic analysis of the integrals (\ref{I.001}) is the insight it yields into current problems of asymptotic combinatorics, specifically related to graphical enumeration.  The work here builds on the systematic asymptotic analysis of (\ref{I.001}) that was carried out in \cite{EM, EMP}. In particular we take as our starting point the following two theorems from this prior work. 

\begin{thm}\cite{EM} \label{EQMSTHM}
There is a constant $T>0$ such that for (complex) $t$ with $\Re(t) \geq 0$ and $|t| < T$ and for $x = n/N$ in a neighborhood, $U$, of $x=1$, one has a uniformly valid asymptotic expansion
\begin{eqnarray}
\label{I.002} \ \ \ \log \tau^2_{n,N}(t) =
n^{2} e_{0}(x, {t}) + e_{1}(x, {{t}}) + \frac{1}{n^{2}}
e_{2}(x, t) + \cdots.
\end{eqnarray}
as $N \to \infty$. 
The meaning of this expansion is:  if you keep terms up to order
$n^{-2h}$, the error term is bounded by $C n^{-2h-2}$, where the
constant $C$ is independent of $x$ and $t$ in the domain $\left\{x\in U; t \in (0 \leq \Re t) \cap ( |t| < T ) \right\}$.  For each $\ell$, the function $e_{\ell}(x, {t})$ is a locally analytic
function of $t$ in a (complex) neighborhood of $t=0$.  Moreover, the asymptotic
expansion of $t$-derivatives of $\log{ \left( Z^{(n)}_{N} \right)}$
may be calculated via term-by-term differentiation of the above
series. $\Box$
\end{thm}

\begin{remark}
In \cite{EM}, the result was stated only for the special case when $x=1$. However, it is straightforward to extend this result to the more general setting stated here and this was explained in \cite{EMP}.  For the later parts of this paper (sections 3 and higher) we will take $x=1$ so that 
$n = N$. In this case we will denote the partition function $Z^{(N)}_{N}$ by just $Z_{N}$ and $e_g(t)$ will denote $e_g(1,t)$. $\Box$
\end{remark}
Our focus will be on the coefficients, $e_g(t)$, which have a direct combinatorial meaning that we now briefly explain.
\smallskip
The $e_g$ enumerate labeled maps. This has already been described in the opening paragraphs of the Introduction. Here we will give the formal and more precise definition.
\medskip
  
A {\it map} $D$ on a compact, oriented and connected surface $X$ is a
pair $D = (K(D), [\imath])$ where
\begin{itemize}
\item $K(D)$ is a connected 1-complex; \item $[\imath]$ is an
isotopical class of inclusions $\imath:K(D) \rightarrow X$; \item
the complement of $K(D)$ in $X$ is a disjoint union of open cells
(faces); \item the complement of the vertices in $K(D)$ is
a disjoint union of open segments (edges).
\end{itemize}
We introduce the notion of a \emph{g-map} which is a map in which the
surface $X$ is the closed, oriented Riemann surface of genus $g$ and
which in addition carries a labeling (ordering) of the vertices.

\begin{thm} \label{thm:01}\cite{EM}
\label{GeomAsymp} The coefficients in the asymptotic expansion
(\ref{I.002}) satisfy the following relations.  Let $g$ be a
nonnegative integer.  Then
\begin{equation} \label{genusexpA}
e_g(t) =  \sum_{j \geq 1}\frac{1}{j!} (-t)^{j}\kappa_g(j)
\end{equation}
in which each of the coefficients $\kappa_g(j)$ is the number of g-maps with $j$  $2\nu$-valent vertices.
\end{thm}

\begin{remark}
Due to this interpretation, the asymptotic expansion (\ref{I.002}) is often referred to as the {\it genus expansion}. 
\end{remark}

\begin{remark}
It is natural to inquire whether there is a generating function representation for {\it unlabeled} maps. After all it is these objects that would seem to have the more purely geometric character. Indeed if it were the case that the action of relabeling a labeled map always produced a distinct other labeled map, then for each unlabeled map with $j$ vertices, one would have $j!$ distinct labeled maps and $\frac{\kappa_g(j)}{j!}$ (the {\it actual} Taylor coefficients of $e_g(t)$ up to a sign) would give the count of unlabeled maps. However, it can happen that a relabeling produces the same labeled map and hence this way of counting is not generally valid. There are some ad hoc fixes for this but we will not go into these since that would take us too far from our main thrust; but, it is worth noting that the kinds of isotropies that cause the problem tend to be restricted to low values of the parameters.  So for certain asymptotic considerations, such as the behavior of these coefficients as $j \to \infty$, the effect of the isotropies "wash out". See Section \ref{other} for a related discussion. 

One may also ask about the consideration of more general weights, for instance of the form 
$$V(\lambda; \ t_{1}, \ldots, t_{\upsilon} ) = \frac{1}{2} \lambda^{2} + \sum_{m=1}^{\upsilon} t_{m}\lambda^{m},$$
which would correspond to allowing for odd valences and, moreover, for {\it mixed} valences  when there are two or more non-vanishing coupling constants, $t_m$. In fact versions of both Theorems \ref{EQMSTHM} and \ref{thm:01} were proven in this generality in \cite{EM}. The results presented in this paper could, in principle, be developed in greater generality along these lines. However, to avoid over-complication of the main ideas we will restrict attention here to the case of pure (but arbitrary), even valence. $\Box$
\end{remark}

\subsection{Different Coupling Weights} \label{diffweights}There is another (discrete) parameter, $\nu$, in the partition function (\ref{I.001}) that determines the exponential weight $V(\lambda)$ in the associated measures. (Graphically it determines the uniform vertex valence of the maps enumerated by the $e_g$.) The associated coupling parameter $t$ in the weight should more properly be denoted by $t_{2\nu}$ to distinguish the different weights. However, as is evident from statements in previously cited theorems of this introduction, we have in general suppressed the dependence on $\nu$ in expressions related to $Z^{(n)}_{N}$ or derived from it.  In part this is to avoid cumbersome notation. In all results and arguments presented in this paper, $\nu$ is a fixed parameter. It does not appear in any of the scalings we discuss. The role that $\nu$ plays should always be clear from context. 

But there is a more salient structural reason why it is natural to omit the explicit $\nu$-dependence:  the formulae for coefficients in the genus expansion that we derive all have a {\it universal}  character in $\nu$; that is to say, the structure of each of these formulae is that of a global meromorphic function of an irrational algebraic function $z_0(s)$ (where $s = -t$).  Although $\nu$ does appear as a discrete parameter in each of these functions, it does not alter that  function's essential meromorphic form. For example, we will show (see  Theorem \ref{egratl}) that, for $g \geq 2$, 
\begin{eqnarray*}
e_g(-s) &=& \frac{(z_0(s) - 1)Q_{d(g)}(z_0(s))}{\left(\nu - (\nu - 1)z_0(s)\right)^{o(g)}},
\end{eqnarray*}
where $Q$ is a polynomial of degree $d(g)$. In particular, the integer-valued functions $d(g)$ and $o(g)$ are independent of $\nu$. 
Moreover, in all cases, the explicit dependence on $\nu$ in such formulae can be completely determined by a finite number of specializations at specific values of  $\nu$ ($\nu$ is a positive integer).  Similar structural properties will be shown to hold for the coefficients of related asymptotic expansions. 
\subsection{The Fundamental Generating Function} \label{FGF} The generating functions that we will describe are not rational in $s$; however, as has already been mentioned, we will show that they are rational functions of a fundamental generating function $z_0(s)$. This breaks the problem of describing these generating functions into two parts. The first is to determine the rational functions as explicitly as possible to the point that one could say something about the associated recursion relation or some other explicit finite description. The main point of the present paper is to make progress on this part of the problem. The second part is to describe $z_0(s)$. Although it is not rational we are actually able, based on \cite{EMP}, to deduce very explicit information about this algebraic function. 

Over the course of this and the next section we will present four different ways of defining $z_0(s)$, all of which play a role in describing and applying the structure of the coefficients of the genus expansion and related geometric expansions. These four different characterizations are
\begin{itemize}
\item[(i)] as a generating function for the generalized Catalan numbers (section \ref{GCN});
\item[(ii)]  as the leading order term in the large $N$ expansion of the recurrence coefficients for monic orthogonal polynomials with respect to the exponential weight $\exp(-N V(\lambda))$ (section \ref{recurrence});
\item[(iii)] as the self-similar solution of an inviscid Burgers equation (section \ref{IBE});
\item[(iv)] in terms of the endpoints of the support for the equilibrium measure of the large $N$ limit of random Hermitian matrices (section \ref{z0-eqmeas}). 
\end{itemize}  

Based on the previous paragraphs, there are several ways in which we can talk about the generating functions in the genus expansion. When we want to think of them directly as functions of the coupling parameter we will write $e_g(t)$ or equivalently $e_g(-s)$ (assuming, when we do this,  that $x=1$). When thinking of its structure as a function of $z_0$ we will write $e_g(z_0)$ or, sometimes, $e_g(z)$. Finally the two perspectives are related by $e_g(-s) = e_g(z_0(s))$. Similar conventions will apply to related expansion coefficients such as the $z_g$ first described in section \ref{recurrence}.

\subsection{Statement of Results}

The first steps toward elucidating the structure of the generating functions $e_g(t)$ were taken in \cite{EMP}. There, differential equations for these functions were derived that made it in principle possible to recursively solve for the $e_g$ in terms of a distinguished combinatorial generating function denoted $z_0(s)$. The differential equations in question will be reviewed in section \ref{background}, but we want first to give a brief overview of the results of this paper and how they depend on prior work. 
\subsubsection{Generalized Catalan Numbers} \label{GCN}We begin with our first description of the fundamental analytic function $z_0(s)$. 

\begin{eqnarray}
\label{z0}  z_0(s) &=& \sum_{j \geq 0} c_\nu^j \zeta_j s^j \,\,\, \mbox{where}\\
\nonumber c_\nu &=& 2\nu \left(\begin{array}{c}
  2\nu-1\\
  \nu-1\\
\end{array}\right) 
= (\nu + 1)  \left(\begin{array}{c}
  2\nu\\
  \nu+1\\
\end{array}\right) \,\,\, \mbox{and}\\
\nonumber \zeta_j &=&   \frac{1}{j}\left(\begin{array}{c}
  \nu j \\
  j-1 \\
\end{array}\right) = \frac{1}{(\nu-1)j+1}\left(\begin{array}{c}
  \nu j \\
  j \\
\end{array}\right).
\end{eqnarray}
(Section \ref{IBE} explains where this expansion comes from.)
When $\nu = 2$, $\zeta_j$ is the $j^{th}$ Catalan number. For general $\nu$ these are the {\it generalized Catalan numbers} which play a role in a wide variety of enumerative combinatorial problems. For a discussion of these applications see \cite{Pierce}.

\subsubsection{Background for the First Main Result} In \cite{EMP} the first few coefficients of (\ref{I.002}) were calculated explicitly:
\begin{eqnarray*}
 e_0(z_0) &=& \eta(z_0-1)(z_0-r) + \frac{1}{2} \log(z_0) \,\,\, \mbox{where} \\
\nonumber && \eta = \frac{(\nu-1)^2}{4\nu(\nu+1)} \,\,\, \mbox{and} \\
\nonumber && r = \frac{3(\nu+1)}{\nu-1};\\
e_1(z_0) &=& -\frac{1}{12} \log\left( \nu - (\nu-1) z_0 \right);\\
e_2(z_0) &=& \frac{(z_0-1)Q(z_0)}{\left(\nu - (\nu - 1)z_0 \right)^{5}} \,\,\, \mbox{where}\\
Q(z_0) &=&
\frac{(\nu-1)}{2880} \left\{ (-\nu^3 + 5 \nu^4 + 8 \nu^5 )  + (-\nu^2 + 41 \nu^3 - 24 \nu^4 - 16 \nu^5 ) z_0 \right.\\
&+&  (44\nu - 89\nu^2 + 54\nu^3 - 17\nu^4 + 8 \nu^5 )z_0^2  + (-12 -12\nu +108\nu^2 - 132 \nu^3 + 48 \nu^4 ) z_0^3 \\
&+& \left.  (-12 + 48 \nu - 72 \nu^2 + 48 \nu^3 - 12 \nu^4) z_0^4 \right\}.
\end{eqnarray*}
\begin{remark}
The expression for $e_2(z_0)$ given in \cite{EMP} contained additional terms involving possible constants of integration. The authors were able to rule out the presence of these other terms for most, but not all, values of $\nu$. However, it is a consequence of the results presented  here (Theorem \ref{egratl}) that these terms are absent for all values of $\nu$ and hence the expression for $e_2(z_0)$ given above is universally valid. $\Box$
\end{remark}
There are several striking features common to these first few coefficients.

First, they are all {\it abelian} functions of the single variable $z_0$. (An abelian function is a function that can be represented as an iterated path integral of a rational function. The rational functions are a minimal subclass of the class of abelian functions.) 

Second, the singularities of these functions are restricted to $z_0 = 0, \nu/(\nu - 1)$. 

Third, the coefficients of these generating functions can be determined by substitution of the generating function for the higher Catalan numbers, $z_0(s)$, into these abelian functions. 

Finally, although the coefficients of these representations do depend (rationally over $\mathbb{Q}$) on $\nu$, they have a universal character; in particular, one should note that knowledge of the coefficients for relatively few values of $\nu$ suffices to completely determine these functions for all values of $\nu$. 

The four above-stated features were established in \cite{EMP} (see Theorem \ref{emp} of this paper) to hold for all the genus expansion coefficients. But we will show that the coefficients, $e_g$, are even more tightly structured than these properties might suggest:

\subsubsection{Statement of the First Main Result} 
\begin{thm} \label{egratl}
For $g \geq 2$ the coefficients of (\ref{I.002}) are rational functions of $z_0$ of the form 
\begin{eqnarray*}
e_g(z_0)  &=& \frac{(z_0-1)Q_{d(g)}(z_0)}{\left(\nu - (\nu - 1)z_0 \right)^{o(g)}}
\end{eqnarray*}
where $Q$ is a polynomial of degree $d(g)$ whose coefficients are rational functions of $\nu$ over the rational numbers $\mathbb{Q}$.  The exponent $o(g)$ and the degree $d(g)$ are non-negative integers to be determined. 
\end{thm}
Once the rational expression for $e_g$ has been determined, the map counts $\kappa_g(n)$ may be read off directly by methods of classical function theory. This will be demonstrated for a related counting problem in Section \ref{counts}.
More could undoubtedly be said about the structure of the $e_g$, but that will be developed elsewhere.
\medskip

\subsubsection{Recurrence Coefficients} \label{recurrence} We turn next to a related map enumeration problem which also has applications to the theory of orthogonal polynomials with exponential weights.  Let $\pi_{n,N}(\lambda)$ denote the $n^{th}$ monic orthogonal polynomial with respect to the exponential weight $\exp(-NV(\lambda))$. These polynomials satisfy a three-term recurrence relation of the form
\begin{eqnarray*}
\pi_{n+1,N}(\lambda) &=& \lambda \pi_{n,N}(\lambda) - b_{n,N}^2(t)  \pi_{n-1,N}(\lambda)
\end{eqnarray*}
Further background on these orthogonal polynomials and their relation to the partition functions (\ref{I.001}) is presented in the Appendix.

In \cite{EMP} it was shown that, as a consequence of Theorem \ref{EQMSTHM},  
\begin{thm} \cite{EMP} \label{II.002}
The recurrence coefficients for the monic orthogonal polynomials with weight $\exp(-N V(\lambda))$ have a full asymptotic expansion, uniformly valid for $t \in (0 \leq \Re t) \cap ( |t| < T )$ and $x = n/N$ in a neighborhood, $U$, of $x=1$, of the form
\begin{eqnarray}\label{z-exp}
b_{n,N}^2(t)  &=&  x (z_0(s)+\frac{1}{n^2}z_1(s)+\frac{1}{n^4}z_2(s)+\cdots)
\end{eqnarray}
where $s = -x^{\nu - 1} t$ (see Section \ref{fs} for the explanation of this scaling) and
\begin{eqnarray} \label{ic}
b_{n,N}^2(0)  &=& n/N = x.
\end{eqnarray} 
\end{thm}
\begin{remark}
This result gives us our second interpretation of $z_0$: it is the leading order term in the large $N$ expansion of the recurrence coefficient $b_{N,N}^2(t)$. The fact that this leading coefficient does indeed coincide with (\ref{z0}) follows from the derivation that will be reviewed in Section
\ref{IBE}.
$\Box$
\end{remark}
The relation of these expansion coefficients to map enumeration is given by the following \cite{EMP}:
\begin{eqnarray} \label{2legs}
 z_g^{(j)}(0) &=& \frac{d^j z_g}{ds^j}|_{s=0} = ^\# \{\mbox{two-legged $g$-maps with $j$ $2\nu$-valent vertices }\}.
\end{eqnarray}
(A \emph{leg} is an edge emerging from a univalent vertex; so that the leg is the only edge incident to that vertex.Thus the maps being counted here have exactly two vertices of valence one; all other vertices have valence $2\nu$.)
\smallskip

\subsubsection{Background for the Second Main Result} In \cite{EMP}, the first few higher order terms in the expansion (\ref{z-exp}) were calculated explicitly:
\begin{eqnarray} 
\label{z1} &&z_1(z_0) =  \frac{  z_0 (z_0 -1) \left\{\frac{(\nu-1) \nu}{12 } \left[(\nu^2+\nu- 2) z_0 -\nu^2 \right]\right\}}{(\nu - (\nu-1)z_0 )^4}\\
\nonumber && = z_0 \left\{\frac{\nu(\nu+2)/12}{\left(\nu - (\nu - 1)z_0 \right)^2} + \frac{-\nu(3\nu+2)/12}{\left(\nu - (\nu - 1)z_0 \right)^3} + \frac{\nu^2/6}{\left(\nu - (\nu - 1)z_0 \right)^4}\right\}\\
\label{z2} && z_2(z_0) =\frac{z_0  (z_0 -1) P_{4}(z_0)}{(\nu -(\nu -1)z_0)^{9}} \,\,\, \mbox{where}\\
\nonumber && P_4(z_0) = \frac{1}{1440} (\nu-1) \nu \left[ (2 \nu^6 -
14 \nu^7 + 24 \nu^8)
\right.  \\
\nonumber &+& (-12 \nu^3 + 148 \nu^4 -546 \nu^5 + 758 \nu^6 - 252
\nu^7 - 96 \nu^8 ) z_0\\
\nonumber &+&  (264 \nu^2 - 1510 \nu^3 + 25551 \nu^4 - 500 \nu^5 -1789
\nu^6 + 840 \nu^7 + 144 \nu^8 ) z_0^2\\
\nonumber &+&  (-536 \nu + 1396 \nu^2 + 912 \nu^3 -4596 \nu^4 + 2492
\nu^5 + 1296 \nu^6 - 868 \nu^7 - 96 \nu^8 ) z_0^3
 \\
\nonumber &+& \left.  (168 + 234 \nu - 1467 \nu^2 + 558 \nu^3 + 1902 \nu^4 -
1446 \nu^5 - 267 \nu^6 + 294 \nu^7 + 24 \nu^8 ) z_0^4 \right]\\
\label{z3} &&z_3(z_0) = \frac{z_0  (z_0 -1) P_{7}(z_0)}{(\nu -(\nu -1)z_0)^{14}} \,\,\, \mbox{where}\\
\nonumber && P_7(z_0) \,\, \mbox{is a polynomial of degree 7 (for the explicit expression, see \cite{EMP}, pp. 66-67)}.
\end{eqnarray}

\subsubsection{Statement of the Second Main Result} We will show that the pattern which appears to be emerging in the preceding examples is in fact the general structure for the coefficients $z_g$. Note that we adopt the same notational conventions for functional dependence of the 
$z_g$ that were described for the $e_g$ at the end of section \ref{FGF}.
\begin{thm} \label{result}
\begin{eqnarray} \label{rational}
\nonumber z_g (z_0) &=& \frac{z_0  (z_0 -1) P_{3g-2}(z_0)}{(\nu -(\nu -1)z_0)^{5g-1}}\\
 &=& z_0 \left\{ \frac{a_0^{(g)}(\nu)}{(\nu - (\nu-1)z_0)^{2g}} + \frac{a_1^{(g)}(\nu)}{(\nu - (\nu-1)z_0)^{2g+1}}+ \cdots + \frac{a_{3g-1}^{(g)}(\nu)}{(\nu - (\nu-1)z_0)^{5g-1}}\right\}, 
\end{eqnarray}
where $P_{3g-2}$ is a polynomial of degree $3g-2$ in $z_0$ whose coefficients are rational functions of $\nu$ over the rational numbers $\mathbb{Q}$. 
\end{thm}
This theorem has a number of remarkable corollaries. 

\subsubsection{Residue Formulae for Graphical Enumeration} For instance, we find that 
\begin{eqnarray*}
&&^\# \{\mbox{two-legged
$g$-maps with $j$ $2\nu$-valent vertices }\}\\   
&=& c_\nu^{j} \sum_{\ell=0}^{3g-1+ j} a_\ell^{(g,j)}(\nu)\,\,\, \mbox{where}\\
a_\ell^{(g,j)}(\nu) &=&  [(j-1)\nu - (2g + \ell + (j-2))] a_\ell^{(g, j-1)}(\nu) + \nu [2g + \ell + (j-2)] a_{\ell - 1}^{(g, j-1)}(\nu) \,\,\, \mbox{and}\\
&& a_\ell^{(g,0)}(\nu) = a_\ell^{(g)}(\nu), \,\,\, \mbox{which is implicitly defined as the $\ell^{th}$ residue in} \,\,\, (\ref{rational}).
\end{eqnarray*}

\subsubsection{Relation with the First Painlev{\'e} Transcendent} When $\nu = 2$ we will show that
\begin{eqnarray*}
a_{3g-1}^{(g)} &=& - 2^{5g-1} \left( 2/3\right)^{g/2} \alpha_g \,\,\, \mbox{for} \,\,\, g \geq 1\\
a_\ell^{(0)} &=& \delta_{\ell,0},
\end{eqnarray*}
where the $\{\alpha_g\}$ are the coefficients in the asymptotic expansion of a class of solutions, (\ref{PI-exp}), to the first Painlev{\'e} equation. As will be discussed in section \ref{dsII}, this points to a broader connection of these enumerative coefficients, for general $\nu$, to the first Painlev{\'e} hierarchy and of the other fundamental residues, $a_\ell^{(g)}(\nu)$, in (\ref{rational}),  to the higher order terms in the expansion of the double scaling limit of the recurrence coefficients.

\subsection{Outline}
In Section \ref{background} we will review the differential equations for the generating functions $z_g(s)$ and also provide additional characterizations of the fundamental generating function $z_0(s)$. In particular we will describe the relation between the polar singularity of $z_g(z_0)$, at $z_0 = \nu/(\nu-1)$, and caustic formation in a Burgers equation for $z_0$. With this background, we give the proof of Theorem \ref{result} in Section \ref{ratl-rec}. The differential equations for the generating functions $e_g$ are reviewed in Section \ref{partfcn-asymp} and then Theorem \ref{egratl} is proved.   Applications of and corollaries to Theorem \ref{result} are presented in Section \ref{applications}. These include those results mentioned at the end of the previous subsection.  Finally, in Section \ref{other} we draw some analogies with and possible connections to other recent work in the theory of map enumeration. 

Appendix A presents the essential elements that we use related to orthogonal polynomials with exponential weights, their recurrence coefficients  and the Toda Lattice equations which describe how these recurrence coefficients transform when the parameters in the weights are varied. We also indicate how the differential equations reviewed in section \ref{background} are related to a continuum limit of the Toda lattice. In Section \ref{forcing-coeffs} a new combinatorial formula is derived for the forcing coefficients that appear in the differential equations for $z_g$. These have relevance for calculating the correlation functions for certain tied random walks on a one-dimensional lattice. 

In Appendix B we briefly review the one-point correlation functions for  eigenvalues of the UE ensembles and the Riemann-Hilbert methods that were used to deduce the structure of their asymptotic expansions. The purpose of this is to give points of reference for some of the theorems quoted in the introduction (particularly Theorems \ref{EQMSTHM} and \ref{II.002}). The Appendix also outlines an extension of these prior results to describe the large \textit{time} behavior of the coefficients $z_g$ and $e_g$.  This extension is needed in the proofs of Theorems \ref{result} and \ref{egratl}.

\section{Background on Differential Equations for Higher Genus Coefficients} \label{background}
A complete differential characterization of the asymptotic coefficients, $z_g$, appearing in (\ref{z-exp}) is given by the first three theorems below. This characterization is founded on a continuum limit of the well-known system of differential equations (the {\it Toda Lattice} equations) which, in this setting, describe how the recurrence coefficients evolve with $t$. In Appendix A the reader will find the pertinent background on Toda equations and a summary of how their continuum limit (\ref{Burgers}) is derived. 

\subsection{Scalings and Notation} The principal results of this paper all involve studying the behavior of certain functions of the partition function in the asymptotic limit as the parameters $N, n$ and/or $t$ tend to infinity (or some critical value), typically with some combination of these parameters being held fixed. In this subsection we summarize the various scalings that will be considered. We will also take this opportunity to state some notational conventions that will be used systematically throughout the paper. 
\subsubsection{Fine Scaling} \label{fs} For several reasons, and in particular when we discuss, in section \ref{background} the derivation of the differential equations which yield explicit expressions for $e_g$ and related generating functions, it will be most effective to use the fine scaling representation of the partition function in terms of the GUE expectation that was given in (\ref{GUEEXPNorm}). There the natural scaled variable to consider was $\frac{t}{N^{\nu-1}}$. There is a closely related fine scaling that will play a role in our second main result related to recurrence coefficients (see section \ref{recurrence}). The scaling variable in this context will be given by $\frac{s}{n^{\nu-1}}$,  where $s$ serves as a similarity variable relating $x$ and $t$ (see (\ref{similarity})). Recall that $n$ and $N$ grow at the same rate so that $x=n/N$ remains fixed at a finite, non-zero value. The relations between all these scalings may be succinctly summarized as follows:
\begin{eqnarray}
\label{finescale} - \frac{1}{2} \frac{s}{n^{\nu-1}} &=& \theta \,\, = \frac{1}{2} \frac{t}{N^{\nu-1}}\\
\label{similarity} s &=& -x^{\nu - 1} t.
\end{eqnarray}
The {\it macroscopic} variable $\theta$ will serve to relate the partition function to standard expressions for {\it tau functions} and {\it Hirota formulae} that play a role in the developments described in sections \ref{recurrence} and \ref{background}. Note that when $x=1, s = -t$. The minus sign between $s$ and $t$ has been introduced because, although the current usage of $t$ is standard for most representations of random matrix partition functions, $s \,\,  (=-t)$ is the natural variable for the generating functions that appear in the genus expansion. See, for instance, formula (\ref{genusexpA}) above. The factors of $\frac{1}{2}$ are introduced in (\ref{finescale}) in order to remove an overall factor of  $\frac{1}{2}$
in the continuum Toda equations (\ref{Burgers}). 

\subsubsection{Spatial Fine Scaling} We will in addition sometimes need to compare asymptotic expansions in $\{n^{-r}\}$ to those in $\{(n+\ell)^{-r}\}$ where $\ell$ is small (and bounded) in comparison to $n$. Of course, these are equivalent asymptotic gauges and it is natural to relate them by a power of the {\it spatial} scaling variable
\begin{eqnarray}
\label{gaugevar} w &=& \left(1 + \frac{\ell}{n}\right)
\end{eqnarray}
which we think of as a quantity close to $1$. This is similar to $x$ but arises in a different context. This variable also serves to mediate between the associated fine scaling variables. For instance if 
\begin{eqnarray}
\frac{s}{n^{\nu-1}} &=& -2\theta = \frac{\tilde{s}}{(n+\ell)^{\nu-1}}, \mbox{then},\\
\tilde{s} &=& w^{\nu-1} s. 
\end{eqnarray}

\subsection{Continuum Limits and Cluster Expansions}
\begin{thm} \label{cont} \cite{EMP} The continuum limit of the Toda Lattice equations (\ref{beqns}) with respect to the time scaling (\ref{finescale}) and the lattice scaling (\ref{gaugevar}) as $n \to \infty$ is given by the following infinite order partial differential equation for $f(s,w)$:
\begin{align}
\nonumber f_s =  F^{(\nu)}\left(n^{-1}; f, f_w, \dots, f_{w^m}, \dots \right) & \doteq  \\
\label{Burgers} c_\nu f^\nu f_w  +
\frac{1}{n^2}F^{(\nu)}_1(f,f_w,f_{ww},f_{www})& +\cdots\\
\nonumber  +
\frac{1}{n^{2g}} F_g^{(\nu)}(f, f_w, f_{w^{(2)}}, \cdots,
f_{w^{(2g+1)}}) & + \cdots \\
\nonumber \textrm{for}\,\, (s,w) \,\, \textrm{near}\,\, (0,1)\,\,  \mbox{and initial data given by}\,\, f(0,w) = w. & \\
\label{contHO}
F_g^{(\nu)}  =  \sum_{\lambda:|\lambda|= 2g+1 \ni \; \ell(\lambda) \leq \nu+1} \frac{d_\lambda^{(\nu,g)}}{\prod_j r_j(\lambda)! } f^{\nu -
\ell(\lambda)+1}\prod_{j} \left( \frac{f_{w^{(j)}}}{j!}
\right)^{r_j(\lambda)} &
\end{align}
where $\lambda= (\lambda_1, \lambda_2, \dots)$ is a
partition, with $\lambda_1 \geq \lambda_2 \geq \lambda_3 \geq \cdots$,  of $2g+1$; $r_j(\lambda) = \# \{\lambda_i | \lambda_i = j\}$; $\ell(\lambda) = \sum_j r_j(\lambda)$ is the {\em length} of $\lambda$; and $|\lambda| = \sum_i \lambda_i$ is the {\em size} of $\lambda$; and 
$d_\lambda^{(\nu,g)}$ are coefficients to be described in the next proposition.
\end{thm}

The following proposition gives an explicit closed form expression for the coefficients $d_\lambda^{(\nu,g)}$. This is a new result. Its proof will be given in \ref{forcing-coeffs}. 

\begin{prop} \label{d-coeffs}
\begin{eqnarray*}
d_\lambda^{(\nu,g)} &=&  \sum_{\begin{array}{c}(\nu+1, \nu, \dots, 2,1) \subseteq \mu \subseteq (2\nu, 2\nu-1, \dots, \nu)\\ \mu \in \mathcal{R}\end{array}} 2 \,\, m_\lambda \left(\mu_1- \eta_1,  \dots, \mu_{\nu+1} - \eta_{\nu+1}\right)
\end{eqnarray*}
where $\mathcal{R}$ is the set of {\it restricted partitions} (meaning that $\mu_1 > \mu_2 > \cdots > \mu_{\nu+1} $), $(\eta_1, \dots, \eta_{\nu+1}) = (2\nu, 2\nu-2, \dots, 2,0)$, and $m_\lambda(x_1, \dots, x_{\nu+1})$ is the monomial symmetric polynomial associated to $\lambda$ \cite{Mac}. The relation of inclusion between partitions, $\rho \subseteq \mu$ means that $\mu_j \geq \rho_j$ for all $j$.
\end{prop}

\begin{remark}
As a polynomial in the $w$-derivatives of $f$, $F_g^{(\nu)}$ is reminiscent of the {\it partial Bell polynomials}, $B_{|\lambda|, \ell(\lambda)}$; however, it is different from these in that it depends on an additional parameter, $\nu + 1$, and the coefficients, which  depend on $d_\lambda^{(\nu,g)}$, are combinatorially more complex. We note also that these coefficients differ slightly from those presented in \cite{EMP} in that the analogous coefficients in the former paper incorporated the parts multiplicities factor, 
$1/\prod_j r_j! $, into the coefficient itself.
\end{remark}

In common parlance, one would take the continuum limit in Theorem \ref{cont} to be the limit of these equations when $n \to \infty$; i.e., the {\it leading order} terms which constitute a finite order (in fact first order) pde. We will in fact be focussing on this leading order equation in the next two subsections. However, for the general analysis we want to carry out in this paper, we will need to look at the successive higher order corrections present in (\ref{Burgers}). In fact we will need to consider these corrections to all orders. So we are interested in the hierarchy of finite order pdes generated by this expansion. This is completely in the spirit of the asymptotic analysis of weakly nonlinear differential equations. As in that setting, we posit the solution to have the form of an asymptotic expansion:
\begin{eqnarray}\label{contexp}
 f(s,w) &=& f_0(s,w) + \frac{1}{n^2}f_1(s,w)+ \cdots + \frac{1}{n^{2g}}f_g(s,w) + \cdots.
\end{eqnarray} 
The hierarchy of equations in question may then be developed by inserting (\ref{contexp}) into (\ref{Burgers}) and then successively, as $g$ increases, collecting terms of order in $n^{-2g}$.  It should be clear that at each order these equations will be partial differential equations, but in several dependent variables: $f_0, f_1, \dots, f_{g-1}, f_g$. It should also be clear that this hierarchy is a triangular system in these dependent variables: at order 
$2g$ the pde is an equation for $f_g(s,w)$ {\it forced} by differential terms in $f_0, f_1, \dots, f_{g-1}$ which should already be known functions having been determined by solving the lower order equations in the hierarchy. These statements and the structure of the hierarchy are made precise in the next theorem.

\begin{thm} \label{empexp} \cite{EMP} The order $2g$ equation, for $g > 0$, in the asymptotic expansion of (\ref{Burgers}) on (\ref{contexp}) is
\begin{eqnarray} \label{II.006}
\frac{d f_g}{ds} &=& c_\nu \left( (f_0)^\nu (f_g)_w + \nu (f_0)^{\nu-1} (f_0)_w f_g \right) + \mbox{\textup{Forcing}}_g,\,\, \mbox{where}\\
\label{Forcing}
\mbox{\textup{Forcing}}_g &=& \left(\frac{c_\nu}{\nu+1}
\frac{\partial}{\partial w} \sum_{\begin{array}{c}
  0 \leq k_j < g \\
  k_1 + \dots + k_{\nu+1} = g\\
\end{array}} f_{k_1}\cdots f_{k_{\nu +1}}\right)\\
\nonumber &+& F_1^{(\nu)}[2g-2] +
F_2^{(\nu)}[2g-4] + \cdots + F_{g}^{(\nu)}[0].
\end{eqnarray}
$F_\ell^{(\nu)}[2r]$ denotes the coefficient of $n^{-2r}$ in $F_\ell^{(\nu)}$. Note, in particular, that such a term cannot involve $f_k$ if $k \geq r$. In (\ref{Forcing}) it is natural, and it will be convenient, to denote the first group of terms by $F_0^{(\nu)}[2g]$.  Also, for $g \ne 0, f_g(0,w) =0$ and, at leading order, $f_0(0,w) = w$.
\end{thm}
The structure of (\ref{II.006}) and (\ref{Forcing}) suggests that these equations may be solved recursively unless a {\it resonance} arises at some stage that obstructs the solution of the linear equation (\ref{II.006}) and requires the introduction of a solvability condition in order to continue. Absent such resonances, one expects to find expressions for each $f_g$, for $g > 0$, as functions, in some appropriate function class, of just $f_0$.  Generally, in the asymptotic analysis of weakly nonlinear differential equations, one does not expect to get away without running into resonances. However, in our special case we do in fact avoid resonances at all stages. This was established in \cite{EMP}. In fact more was shown there: the formal series (\ref{contexp}) that can be built from these recursively constructed coefficients is in fact an asymptotic expansion uniformly valid for $s$ negative and near $0$.  The precise statements are given in the next theorem. Appendix \ref{continuum} outlines the essential elements underlying these results.

\begin{thm}\label{emp} \cite{EMP}

\begin{enumerate}
\item[(i)] 
\begin{eqnarray}
\label{ssscale} f_g(s,w)&=& w^{1-2g}z_g(w^{\nu-1}s)\\
\nonumber f(s,1) &=& z_0(s)+\frac{1}{n^2}z_1(s)+\frac{1}{n^4}z_2(s)+\cdots .
\end{eqnarray}

\item[(ii)] The coefficient $z_{g}$  is
an abelian function of $z_{0}$ with singularities only possible at $z_0 =
0$ and $z_0=\nu/(\nu-1)$.

\item[(iii)]
The coefficient $z_{g}$ is more explicitly presented as a function of
$z_{0}$ through the following integral equation:
\begin{equation*}
z_g(s)= z_g(z_0(s)) = \frac{z_0(s)^{2(1-g)}}{\nu-(\nu-1)z_0(s)} \int_1^{z_0(s)}
\frac{(\nu-(\nu-1)z)}{c_\nu z^{\nu+3-2g} }
\mbox{\textup{Forcing}}_g(z) dz,
\end{equation*}
where 
\begin{eqnarray*}
\mbox{\textup{Forcing}}_g(z_0) &=& \left. \mbox{\textup{Forcing}}_g\right|_{w=1}.
\end{eqnarray*}
The terms in $\mbox{\textup{Forcing}}_g|_{w=1}$ depend only on $z_k(z_0), k<g$ and their derivatives and these derivatives can in turn be re-expressed in terms of $z_k(z_0)$ for $k<g$. (See sections \ref{Fstruc} and \ref{Lstruc}).
\end{enumerate}
\end{thm}

Note that $b_{n,N}^{2}$ (see (\ref{z-exp})) and $x f(s,1)$ possess the same asymptotic expansion. Note also that since $f(0,w) = w$, and consistent with (\ref{ic}), 
\begin{eqnarray*}
z_0(0) &=& 1\\
z_g(0) &=& 0 \,\,\, \mbox{for}\,\,\,  g \geq 1.
\end{eqnarray*}
A principal result of this paper is to refine part (ii) of the previous theorem to state that $z_g$ is a {\it rational} function of $z_0$ for $g > 1$. The rational functions are a subclass of the abelian functions. Another way to say this is that we establish solvability, without resonances, of the differential hierarchy (\ref{II.006}), beyond $g = 1$,  within the function class of rational functions. There is a solvability condition at $g=1$ which is to choose a branch of the logarithm at $z_0 = \nu/(\nu-1)$.

\subsection{Leading Order} \label{IBE}
The leading equation in the PDE scheme (\ref{Burgers}) is
\begin{eqnarray} \label{Burgers0}
f_s &=&  c_\nu f^\nu f_w\\
\nonumber f(0,w) &=& w,
\end{eqnarray}
which is an instance of the {\it inviscid Burgers equation}, well-known from the theory of shock waves. At leading order in $n$ this becomes
\begin{eqnarray*}
\frac{d f_0}{ds} &=& c_\nu (f_0)^\nu (f_0)_w.
\end{eqnarray*}
Substituting the self-similar scaling (\ref{ssscale}) and then evaluating at $w=1$ it further reduces to the ODE
\begin{eqnarray} \label{z0-de}
z_0'(s) = c_\nu z_0(s)^\nu \left( z_0(s) + (\nu-1) s
z_0'(s)\right)
\end{eqnarray}
with initial condition $z_0(0) = 1$. The solution is given implicitly by 
\begin{equation} \label{II.005}
1 = z_0(s) - c_\nu s z_0(s)^\nu,
\end{equation} 
which is our third interpretation of $z_0$ and the one from which our first one, (\ref{z0}), was derived. 

\begin{remark}
We mention that, in the case $\nu = 2$, (\ref{II.005}) has a relation to the following discrete equation that played a key role in the early studies of 2D quantum gravity in the physics literature.
\begin{eqnarray} \label{string}
4 t  b_{n,N}^2(t) \left(b_{n-1,N}^2(t) + b_{n,N}^2(t) + b_{n+1,N}^2(t) \right) + b_{n,N}^2(t) = \frac{n}{N}. 
\end{eqnarray}
In the literature on orthogonal polynomials, this relation is one of the basic examples of what are known as {\it Freud equations} \cite{Mag}. 
In the physics literature it was referred to as the {\it string equation}.  It is an example of a discrete Painlev{\'e} equation and, in an appropriate sense, is a discretization of the ordinary differential equation that defines the (continuous) first Painlev{\'e} transcendent \cite{FIKI, FIKII}.  With respect to the continuum limit described in the previous subsection and based on Section \ref{continuum}, (\ref{string}) limits at leading order to (\ref{II.005}) evaluated at $\nu = 2$.
\end{remark}

\subsection{The Burgers Caustic}
\begin{figure} [h]  
   \centering
   \includegraphics[width=2in]{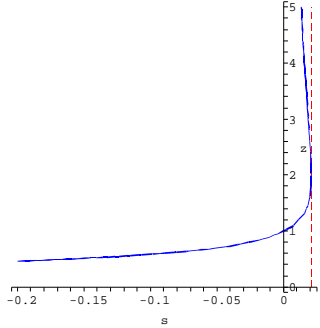} 
   \caption{Burgers Caustic}
   \label{fig:caustic}
\end{figure}

As illustrated in Figure \ref{fig:caustic}, the solution, $z_0$ develops a {\it caustic} where $z_0^\prime$ becomes undefined. (This infinite slope corresponds to the point, in unscaled variables, where characteristics for (\ref{Burgers0}) would first focus or {\it shock}.) 

From (\ref{z0-de}), and using
 (\ref{II.005}),  one can express  the derivative purely in terms of $z_0$ as
\begin{eqnarray} \label{z0-derivative}
z_0^\prime &=& \frac{c_\nu z_0^{\nu + 1}}{\nu - (\nu-1)z_0}.
\end{eqnarray}
From this one sees that the shock coincides with the proposed unique pole location, 
$$
\nu - (\nu-1)z_0 = 0,
$$
in the rational expressions for the higher generating functions. For future use we want to calculate here the {\it shock time} $s_c$ which corresponds to $z_0 = \nu/(\nu-1)$;
\begin{eqnarray} \label{caustic}
\nonumber s_c &=& s\left( z_0 = \frac{\nu}{\nu-1}\right)\\
\nonumber &=& \frac{z_0-1}{c_\nu z_0^\nu}|_{z_0 = \frac{\nu}{\nu-1}}\\
&=& \frac{(\nu-1)^{\nu-1}}{c_\nu \nu^\nu}.
\end{eqnarray}
From (\ref{z0-derivative}) it is straightforward to derive the leading local behavior near the caustic,
\begin{eqnarray} \label{caustic2}
\left( z_0 - \frac{\nu}{\nu-1}\right) &=& - \sqrt{\frac{-2 c_\nu \nu^{\nu+1}}{(\nu - 1)^\nu}(s - s_c)} \left\{1 + \mathcal{O}(s - s_c)\right\}.
\end{eqnarray}
In the special case of $\nu = 2$, $s_c = \frac{1}{48}$ and one can solve globally for $z_0$ from (\ref{II.005}) using the quadratic formula,
\begin{equation*}
z_0 = \frac{1- \sqrt{1 - 48s}}{24s}.
\end{equation*}
\begin{remark}
It follows from these observations and Theorems \ref{egratl} and \ref{result} that the maximal {\it domain of holomorphy}, containing $t=0$, for both $e_g$ and $z_g$ is the set of $t \in \mathbf{C} \backslash (-s_c , \infty)$. 
\end{remark}

\subsection{Another Interpretation of $z_0$} \label{z0-eqmeas}
The fourth and final interpretation of $z_0$ is in terms of the spectral density for random Hermitean matrices with respect to the probability distribution (\ref{RMT}). For the class of weights, $V_\nu$, we are considering and for all $t \geq 0$ (see section \ref{uniform}), the mean density of eigenvalues 
for this ensemble limits to a continuous density supported on a finite interval $[-\beta, \beta]$. This density defines a probability measure, usually referred to as the {\it equilibrium measure}, which has the following characterization.

\begin{thm} \cite{DKM}
The equilibrium measure $\mu_V$  is absolutely continuous with respect to Lebesgue measure, and
\begin{align}
\nonumber d\mu_{V} &= \psi \, d\lambda ,\\
\label{eqmeas} \psi (\lambda ) &= \frac{1}{2 \pi} \chi_{(-\beta ,\beta )} (\lambda)
\sqrt{(\lambda + \beta )(\beta -\lambda )}\, h(\lambda ),
 \end{align}
where $h(\lambda )$ is a polynomial of degree $2\nu -2$, which is
strictly positive on the interval $[-\beta,\beta]$.
\end{thm}
The endpoints of support $\pm \beta(t)$ vary smoothly with $t$ (or equivalently $s$) and
\begin{eqnarray}\label{beta}
z_0(s) &=& \frac{\beta^2(-s)}{4}.
\end{eqnarray} 
The fact that this characterization of $z_0$ coincides with the other definitions given in this and the previous section, follows from the requirement that the equilibrium measure should be a probability measure. This is an algebraic condition on the endpoints of support that is equivalent to
$(\ref{II.005})$. A detailed explanation of this relation is given in \cite{EMP}.

The equilibrium measure is the unique solution of the following variational problem for logarithmic potentials in the presence of an external field
\cite{ST}. (In this setting, $V$ is the external field.) 
\begin{eqnarray} \label{VProb}
&& \sup_{\mu\in \mathbf{A}} \left\{-\int V(\lambda) d\mu(\lambda) + \int \int \log|\lambda-\eta| d\mu(\lambda) d\mu(\eta)\right\},
\end{eqnarray}
where $\mathbf{A}$ is the set of all positive Borel measures on the real axis with unit mass.
This variational problem, and hence the definition of $\mu_V$,  can be extended to $t \in (-s_c, 0)$ \cite{DK}. It follows that this interpretation of $z_0$ extends as well to this negative interval and coincides, there, with its other interpretations given in this section.

\section{Rationality of the Asymptotic Recurrence Coefficients} \label{ratl-rec}

The first step toward proving Theorem \ref{result} is to show that $z_g$ must be a rational function of $z_0$. Given parts (ii) and (iii) of Theorem \ref{emp}, it suffices to show that the integrand in Theorem \ref{emp} (iii) has no residues at $z=0$ and $z=\nu/(\nu - 1)$.    In \cite{EMP} 
this vanishing of residues was established for $z_1,  z_2$ and $z_3$ (in fact, Theorem \ref{result} was fully established in these three cases by direct calculation).  If there is a term of $\mbox{\textup{Forcing}}_g$ which generates a non-zero residue in the integrand of (2) we call that term  {\it resonant}. We will show by induction that for all $g$, $\mbox{Forcing}_g$ can be decomposed as a sum of terms each of which is non-resonant. 

\begin{prop} \label{vanres}
\begin{enumerate}
\item[(i)] $z_g$ is regular at $z_0 = 0$ and in fact vanishes at least simply there.

\item[(ii)] $\mbox{Forcing}_g$ may be written as a sum of terms each of which has the form 
\begin{equation*}
f_0^{\nu - m + 1} (f_{k_1})_{w^{(j_1)}} (f_{k_2})_{w^{(j_2)}} \cdots (f_{k_{m}})_{w^{(j_{m})}}
\end{equation*}
where $0< m \le \nu + 1$ and $ j_1 + \cdots + j_{m} = 2(g - k_1 - k_2 - \cdots - k_{m}) + 1$, with $k_i < g$.

\item[(iii)] For $(k,n) \ne (0,0)$, and assuming that Theorem \ref{result} holds for all $0 < k < g$, 
\begin{eqnarray*}
(f_k)_{w^{(n)}}(s,1) &=& 
z_0\left\{\frac{b_0^{(k,n)}}{(\nu - (\nu-1)z_0)^{2k+n}} + \cdots  +  \frac{b_{3k+n-1}^{(k,n)}}{(\nu - (\nu-1)z_0)^{5k+2n-1}}\right\}. 
\end{eqnarray*}
\end{enumerate}

\end{prop}
\medskip

The proof of the rational form (\ref{rational}) for $z_g$  follows fairly straightforwardly from this proposition. 
\smallskip

 {\bf Proof (of Theorem \ref{result}).} Inserting expansions of the form (iii),  from Proposition \ref{vanres}, for each factor of the product appearing in (ii),  from Proposition \ref{vanres}, one derives the rational form (as a function of $z_0$) of each term in $\mbox{Forcing}_g$ (see Section \ref{Fstruc} below for further explanation).  One may then deduce, from Theorem \ref{emp}(iii), that $z_g$ is a sum of terms, with $K=k_1 + k_2 + \cdots + k_m$ and $J=j_1 + j_2 + \cdots + j_m$, of the form 
\begin{eqnarray}  \label{laurent}
\nonumber && \frac{z_0^{2-2g}}{\nu-(\nu-1)z_0} \int_1^{z_0}
\frac{(\nu-(\nu-1)z)}{c_\nu z^{\nu+3-2g}} \cdot z^{\nu+1} \left\{\frac{D_0}{(\nu - (\nu-1)z)^{2g+1}} + \cdots \right.\\
\nonumber &\dots& +  \left. \frac{D_{3K+J-m-1}}{(\nu - (\nu-1)z)^{5K+2J-m}}\right\} dz\\
\nonumber &=& \frac{z_0^{2-2g}}{c_\nu (\nu-(\nu-1)z_0)} \int_1^{z_0}
 z^{2g-2}\left\{\frac{D_0}{(\nu - (\nu-1)z)^{2g}} + \cdots  \right.\\
 \nonumber &\dots& + \left.  \frac{D_{3K+J-m-1}}{(\nu - (\nu-1)z)^{5K+2J-m-1}}\right\} dz\\
 \nonumber &=& \frac{z_0^{2-2g}}{c_\nu (\nu-(\nu-1)z_0)} \int_1^{z_0}
 \left(\frac{\nu - (\nu -  (\nu -1)z))}{\nu-1}\right)^{2g-2}\left\{\frac{D_0}{(\nu - (\nu-1)z)^{2g}} + \cdots \right.\\
\nonumber  &\dots& + \left.  \frac{D_{3K+J-m-1}}{(\nu - (\nu-1)z)^{5K+2J-m-1}}\right\} dz\\
 \nonumber &=&  \frac{z_0^{2-2g}}{c_\nu (\nu-(\nu-1)z_0)} \int_1^{z_0}
 \left\{\frac{E_0}{(\nu - (\nu-1)z)^{2}} + \cdots  +  \frac{E_{5K+2J-m-3}}{(\nu - (\nu-1)z)^{5K+2J-m-1}}\right\} dz\\
 &=& \frac{1}{z_0^{2g-2}}
 \left\{\frac{\hat{E}_{-1}}{(\nu - (\nu-1)z_0)}+\frac{\hat{E}_0}{(\nu - (\nu-1)z_0)^{2}} + \cdots  +  \frac{\hat{E}_{5K+2J-m-3}}{(\nu - (\nu-1)z_0)^{5K+2J-m-1}}\right\}
\end{eqnarray}
where the leading pole term, with coefficient $\hat{E}_{-1}$, arises from the lower bound of integration which insures that this expression vanishes at $z_0 = 1$. The integrand in the penultimate line above is clearly non-resonant and it therefore follows inductively that $z_g$ is rational as a function of $z_0$. However, in order to establish the form for $z_g$ stated in equation (\ref{rational}) and to complete the induction we need to make further use of this result. The total expression for $z_g$ is comprised of a sum of series of the form (\ref{laurent}), each series coming from 
one of the terms into which $\mbox{Forcing}_g$ may be decomposed (see Proposition \ref{vanres}(ii)).  

We determine next the maximal pole order of this total expression. The pole order of each component series as given by the formula above, is $5K + 2J -m-1$. By Proposition \ref{vanres}: (ii), there is one constraint,  $J = 2(g - K) + 1$, and two boundary conditions, $0 < m \leq \nu + 1$ and $0 \leq K \leq g$. It is straightforward to check that this linear problem is maximized over integer values by $(K,m) = (g,2)$ or $(K,m) = (g-1,1)$, so that, respectively, $J=1$ or $J=3$. Hence the maximal pole order can be at most  $5g-1$. 

When $J=1$, the terms that contribute to the maximal pole are of the form

\begin{equation} \label{maxpole}
\nu c_\nu f_0^{\nu-1} \sum_{m=1}^{g-1}  f_{m}\left( f_{g-m}\right)_w
\end{equation}
as can be seen from (\ref{Burgers}).

When $J=3$, a term yielding this maximal pole order could only come from the $F_1^{(\nu)}[2g-2]$ component of $\mbox{Forcing}_g$ as can be seen from (\ref{Forcing}). Moreover, since $m=1$, the only  term within this component giving this maximal order comes from the summand associated to the partition $\lambda = (3)$. 

More will be said later about the structure of the coefficient of the highest order pole (Section \ref{DS}). For now it will suffice to conclude that 
$z_g$ has the form
\begin{eqnarray} \label{zgexp}
z_g = \frac{1}{z_0^{2g-2}}\left\{\frac{C_{0}}{(\nu - (\nu-1)z_0)}+\frac{C_1}{(\nu - (\nu-1)z_0)^{2}} + \cdots  +  \frac{C_{5g-2}}{(\nu - (\nu-1)z_0)^{5g-1}}\right\}
\end{eqnarray}
for all $\nu \geq 2$. 

By Proposition \ref{vanres}(i), $z_g$ vanishes at $z_0 = 0$. For this to be possible, the Laurent series in (\ref{zgexp}) must have a factor of $z_0^{2g-1}$ and hence this expression for $z_g$ becomes
 \begin{eqnarray} 
\nonumber z_g &=& z_0 \left\{ \frac{a_0^{(g)}}{(\nu - (\nu-1)z_0)^{2g}} + \frac{a_1^{(g)}}{(\nu - (\nu-1)z_0)^{2g+1}}+ \cdots + \frac{a_{3g-1}^{(g)}}{(\nu - (\nu-1)z_0)^{5g-1}}\right\}\\
\label{pole-exp} &=& \frac{z_0  (z_0 -1) P_{3g-2}(z_0)}{(\nu -(\nu -1)z_0)^{5g-1}}
\end{eqnarray}
where the factor of $z_0-1$ in the last line is inferred from the fact that each term contributing to the final expression vanishes at $z_0 = 1$.  This completes the induction and the proof of Theorem \ref{result}. $\Box$

\subsection{Proof of Proposition \ref{vanres} } \label{proofs}

\subsubsection{Regularity at $z_0 = 0$} \label{z0-reg}
Recall from (\ref{z0-derivative}) that $z_0(s)$ satisfies the first order ode
\begin{eqnarray*} 
z_0^\prime = \frac{c_\nu z_0^{\nu+1}}{(\nu - (\nu-1)z_0)}.
\end{eqnarray*}
It follows from this dynamical representation that if $z_0$ is initialized at $1$, then its trajectory, as $s \to -\infty$,  limits monotonically and asymptotically to the fixed point at $z_0 = 0$. Hence, to study the behavior of $z_g$ as $z_0 \to 0$, it suffices to consider 
the limit of $z_g(s)$ as  $s \to -\infty$. By Theorem \ref{extension} one has an asymptotic expansion for the orthogonal polynomial recurrence coefficients that is uniformly valid for $s$ in a neighborhood of the half-line $(-\infty,0)$ in the complex $s$-plane, so that 
\begin{eqnarray} \label{z-asymps}
|b^2_{N,N}(-s)  -z_0(s) -N^{-2} z_1(s) - \cdots -N^{-2g}z_{g}(s)| &<& \frac{C}{N^{2g+2}}  
\end{eqnarray}
where $C$ is independent of $s$ in a neighborhood of the negative $s$-axis.  We proceed inductively and so can assume that $z_j$ for $j <  g$ vanishes as $s \to -\infty$ (we know, to begin with, that $z_0$ vanishes in this limit). If one restricts to $s$ on the negative real half-line, then 
(\ref{z-asymps}) may be re-expressed as
\begin{eqnarray}
\label{ineq_1} z_g - \frac{C}{N^2} &<& N^{2g}\left(b_{N,N}^2(-s) - z_0(s) - N^{-2} z_1(s) - \cdots - N^{-2g+2}z_{g-1}(s)\right)\,\, \mbox{and}\\
\label{ineq_2} z_g  + \frac{C}{N^2} &>&  N^{2g}\left(b_{N,N}^2(-s) - z_0(s) - N^{-2} z_1(s) - \cdots -N^{-2g+2}z_{g-1}(s)\right).
\end{eqnarray}
We now choose and {\it fix} $N$ so that $\frac{C}{N^2} < \frac{\epsilon}{2}$.    Next we note that  
\begin{eqnarray*}
b_{N,N}^2(-s) &\leq& \frac{1}{N} \int_{-\infty}^\infty \lambda^2 p_N^2(s, \lambda) e^{-NV(s,\lambda)} d\lambda\\
 &\to& 0 \,\,\, \mbox{as} \,\,\, s \to -\infty, 
\end{eqnarray*}
where $p_n$ is the $n^{th}$ orthonormal polynomial with respect to the exponential weight $\exp\left(-NV(s,\lambda)\right) = \exp\left(N(\frac{1}{2}\lambda^2 - s \lambda^{2\nu})\right)$. The inequality is a direct consequence of the recursion formula for these orthonormal polynomials. The limit
in the second line follows because the family of densities $\{p_N^2(s, \lambda) e^{-NV(s,\lambda)}\}$ constitutes a delta sequence as 
$s \to - \infty$.

As a consequence of this observation and the induction, one may choose $|s|$ sufficiently large so that the right-hand side of (\ref{ineq_1}) is less than $\epsilon/2$ and the right-hand side of (\ref{ineq_2}) is greater than $-\epsilon/2$. 
Therefore,
\begin{eqnarray*}
|z_g(s)| &<& \frac{C}{N^2} + \frac{\epsilon}{2} < \epsilon
\end{eqnarray*}
for $s$ sufficiently negative.  Since $\epsilon$ was arbitrary, this establishes statement (i) of Proposition \ref{vanres}. 

\subsubsection{The Structure of $\mbox{Forcing}_g$} \label{Fstruc}
By (\ref{Forcing}), $\mbox{Forcing}_g$ is a graded sum
\begin{eqnarray*}
&& F_0^{(\nu)}[2g] + F_1^{(\nu)}[2g-2] + \cdots + F_g[0] 
\end{eqnarray*}
where $F_r^{(\nu)}[2g-2r]$ is the coefficient of $n^{-2g+2r}$ in $F_r^{(\nu)}$.  The terms in $F_r^{(\nu)}$ are in $1-1$ correspondence with the partitions of $2r+1$ as described in (\ref{contHO}); i.e., the terms in this component are each proportional to an expression of the form \begin{equation} \label{Fatoms}
f_0^{\nu - m + 1} (f_{k_1})_{w^{(j_1)}} (f_{k_2})_{w^{(j_2)}} \cdots (f_{k_{m}})_{w^{(j_{m})}}
\end{equation}
with $0 \leq k_i < g$ and where $\sum_{i=1}^m j_i = 2r+1$. The further requirement that these terms are coefficients of the power  
$n^{-2g+2r}$ in $F_r^{(\nu)}$ means that $\sum_{i=1}^m k_i = 2g - 2r$.  Eliminating $r$ from these two equations we may conclude 
that each term of $\mbox{Forcing}_g$ is proportional to an expression of the form (\ref{Fatoms}) where
$ j_1 + \cdots + j_{m} = 2(g - k_1 - k_2 - \cdots - k_{m}) + 1$. This establishes statement (ii) of Proposition \ref{vanres}.

\subsubsection{Laurent Expansions of $\mbox{Forcing}_g$} \label{Lstruc}
In this section we set up and prove the main lemma needed to establish the pole expansions, Proposition \ref{vanres} (iii), of the fundamental factors, $(f_k)_{w^{(n)}}(s,1)$. From (\ref{ssscale}) one has
\begin{equation*}
f_k(s,w) = w^{1-2k}z_k(w^{\nu-1}s).
\end{equation*}
One can calculate the first couple derivatives directly:
\begin{eqnarray*}
(f_{k})_w &=& (1-2k)w^{-2k}z_k + (\nu-1)w^{\nu-1-2k}sz_k^\prime \\
(f_{k})_{ww} &=& -2k(1-2k)w^{-(2k+1)}z_k + (\nu-1)(\nu-4k)w^{\nu-2-2k}sz_k^\prime + (\nu-1)^2 w^{2\nu-3-2k}s^2 z_k^{\prime\prime}
\end{eqnarray*}
The general form of the higher order $w$-derivatives is given by
\begin{eqnarray*}
(f_k)_{w^{(n)}}(s,w) &=& \sum_{j=0}^n (\nu-1)^j P_j^{(n,k)} (\nu) w^{(\nu-1)j - (2k+n-1)}s^j z_k^{(j)}
\end{eqnarray*}
Note that the coefficient of $w^{(\nu-1)j - (2k+n-1)}s^j z_k^{(j)}$ has two sources in $(f_k)_{w^{(n-1)}}$.  These two sources are 
\begin{eqnarray*}
&& (\nu-1)^{j-1} P_{j-1}^{(n-1,k)} (\nu) w^{(\nu-1)(j-1) - (2k+n-1)+1}s^{j-1} z_k^{(j-1)}\\
\mbox{and}&& (\nu-1)^j P_j^{(n-1,k)} (\nu) w^{(\nu-1)j - (2k+n-1)+1}s^j z_k^{(j)}.
\end{eqnarray*}
Given this, one can {\it read off} a recursion formula for the coefficients $P_j^{(n,k)} (\nu)$ that is given in the next lemma.

\begin{lemma}
\begin{eqnarray} \label{P-rec}
(f_k)_{w^{(n)}}(s,1) &=&
\sum_{j=0}^n (\nu-1)^j P_j^{(n,k)} (\nu) s^j z_k^{(j)}
\end{eqnarray}
where
\begin{eqnarray}
P_j^{(n,k)} (\nu) &=&   P_{j-1}^{(n-1,k)} (\nu) + \{ (\nu - 1) j - (2k-2+n) \} P_j^{(n-1,k)} (\nu)\\
\nonumber &\mbox{with}& \\
\label{P-recI} P_0^{(n,k)} (\nu) &=& (1-2k)(-2k)\cdots (1-2k-n+1) \\
\label{P-recII} P_n^{(n,k)} (\nu) &=& 1 \,\,\,\,\, \mbox{and}\\
\label{P-recIII} P_j^{(n,k)} (\nu) &=& 0  \,\,\,\,\, \mbox{for}\,\,\, j > n \,\,\, \mbox{and}\,\,\, j < 0..
\end{eqnarray}
\end{lemma}
We next turn to studying the Laurent expansion of $z_k$ around $z_0 = \frac{\nu}{\nu - 1}$.  This analysis is subordinate to the inductive assumption that for $0 < k < g$, 
\begin{eqnarray*}
z_k (z_0) = \frac{z_0  (z_0 -1) P_{3k-2}(z_0)}{(\nu -(\nu -1)z_0)^{5k-1}}.
\end{eqnarray*}
It follows that
\begin{eqnarray*}
z_k &=& z_0 \left\{ \frac{a_0^{(k,0)}(\nu)}{(\nu - (\nu-1)z_0)^{2k}} + \frac{a_1^{(k,0)}(\nu)}{(\nu - (\nu-1)z_0)^{2k+1}}+ \cdots + \frac{a_{3k-1}^{(k,0)}(\nu)}{(\nu - (\nu-1)z_0)^{5k-1}}\right\}.
\end{eqnarray*} \label{zk-der}
Moreover, differentiating this expression  $j$ times, with respect to $s$, one arrives at
\begin{eqnarray} \label{zkder-pole}
 z_k^{(j)} &=& c_\nu^{j} z_0^{j\nu + 1} \left(\sum_{\ell=0}^{3k-1+ j}\frac{a_\ell^{(k,j)}}{(\nu - (\nu-1)z_0)^{2k+\ell+j}}\right).
 \end{eqnarray}
 To see this, note that
 \begin{eqnarray*}
z_k^{(j)}&=& \frac{dz_k^{(j-1)}}{ds}\\
 &=& c_\nu^{j-1} z_0^{(j-1)\nu}z_0^\prime \left(\sum_{\ell=0}^{3k-1+j-1}\frac{[(j-1)\nu + 1]a_\ell^{(k,j-1)}}{(\nu - (\nu-1)z_0)^{2k+\ell+j-1}}\right. \\
  &+& \left.  (\nu-1)z_0\sum_{\ell=0}^{3k-1+j-1}\frac{(2k+\ell+(j-1))a_\ell^{(k,j-1)}}{(\nu - (\nu-1)z_0)^{2k+\ell+j}}\right)\\
 &=& c_\nu^{j-1} z_0^{(j-1)\nu}z_0^\prime \left(\sum_{\ell=0}^{3k+j-2}\frac{[(j-1)(\nu-1) - 2k - \ell + 1]a_\ell^{(k,j-1)}}{(\nu - (\nu-1)z_0)^{2k+\ell+j-1}} \right. \\
 &+& \left. \nu\sum_{\ell=0}^{3k+j-2}\frac{(2k+\ell+(j-1))a_\ell^{(k,j-1)}}{(\nu - (\nu-1)z_0)^{2k+\ell+j}}\right)\\
 &=& c_\nu^{j} z_0^{j\nu+1}\left(\sum_{\ell=0}^{3k+j-2}\frac{[(j-1)(\nu-1) - 2k - \ell + 1]a_\ell^{(k,j-1)}}{(\nu - (\nu-1)z_0)^{2k+\ell+j}} \right. \\
 &+& \left. \nu\sum_{\ell=0}^{3k+j-2}\frac{(2k+\ell+(j-1))a_\ell^{(k,j-1)}}{(\nu - (\nu-1)z_0)^{2k+\ell+j+1}}\right)\\
 &=& c_\nu^{j} z_0^{j\nu+1}\left(\sum_{\ell=0}^{3k-1+j}\frac{[(j-1)\nu - (2k + \ell + (j-2)] a_\ell^{(k,j-1)} + \nu (2k+\ell+(j-2))a_{\ell-1}^{(k,j-1)}}{(\nu - (\nu-1)z_0)^{2k+\ell+j}} \right)
\end{eqnarray*}
where, in the fourth equation,  the identity (\ref{z0-derivative}), $z_0^\prime = \frac{dz_0}{ds}=\frac{c_\nu z_0^{\nu+1}}{\nu - (\nu-1)z_0}$, was applied. From this one deduces the following recursion for the succession of derived Laurent coefficients
\begin{lemma} \label{L-coeffs}
\begin{eqnarray*}
a_\ell^{(k,j)} &=&  [(j-1)\nu - (2k + \ell + (j-2))] a_\ell^{(k, j-1)} + \nu [2k + \ell + (j-2)] a_{\ell - 1}^{(k, j-1)}
\end{eqnarray*}
with 
\begin{eqnarray}
\nonumber a_0^{(0,0)}&=&1\\
\nonumber a_\ell^{(0,0)} &=& 0 \,\,\,\,\, \mbox{for} \,\,\,\,\, \ell > 0\\
\label{a-vanish} a_\ell^{(k,j)} &=& 0  \,\,\,\,\, \mbox{for} \,\,\,\,\, \ell < 0\,\,\, \mbox{and} \,\,\, \ell \geq 3k+j.
\end{eqnarray}
\end{lemma}
Finally we consider the form of the expressions $s^j z_k^{(j)}$ which appear in  (\ref{P-rec}). 
Recall, from (\ref{II.005}), that
\begin{eqnarray} \label{ess}
s &=& \frac{z_0-1}{c_\nu z_0^\nu},
\end{eqnarray}
so one may deduce from (\ref{zk-der}) that
\begin{eqnarray*}
s^j z_k^{(j)} &=& z_0 (z_0 - 1)^j \sum_{\ell = 0}^{3k-1+j} \frac{a_\ell^{(k,j)}}{(\nu - (\nu-1)z_0)^{2k+\ell+j}} \\
&=& z_0  \frac{(1-(\nu - (\nu-1)z_0))^j}{(\nu-1)^j} \sum_{\ell = 0}^{3k-1+j} \frac{a_\ell^{(k,j)}}{(\nu - (\nu-1)z_0)^{2k+\ell+j}} \\
&=&  \frac{z_0}{(\nu- 1)^j}  \sum_{r=0}^j\sum_{\ell = 0}^{3k-1+j}  \frac{(-1)^{j-r} 
\left(
\begin{array}{c}
 j  \\
j - r  
\end{array}
\right)
 a_\ell^{(k,j)}}{(\nu - (\nu-1)z_0)^{2k+\ell+r}}\\
 &=&  \frac{z_0}{(\nu- 1)^j} \sum_{r=0}^j \sum_{m = r}^{3k+j + r -1}  \frac{(-1)^{j-r} 
\left(
\begin{array}{c}
 j  \\
 r  
\end{array}
\right)
 a_{m-r}^{(k,j)}}{(\nu - (\nu-1)z_0)^{2k+m}}
\end{eqnarray*}
where $m=\ell + r$.  Inserting this last expansion into the expansion for $(f_k)_{w^{(n)}}(s,1)$ given by (\ref{P-rec}), and applying the vanishing conditions (\ref{a-vanish}) to extend the bounds of the inner summation, yields
\begin{eqnarray}
\nonumber (f_k)_{w^{(n)}}(s,1) &=& z_0 \sum_{j=0}^n P_j^{(n,k)} \sum_{r=0}^j\sum_{m = 0}^{3k+2j-1}  \frac{(-1)^{j-r} 
\left(
\begin{array}{c}
 j  \\
 r  
\end{array}
\right)
 a_{m-r}^{(k,j)}}{(\nu - (\nu-1)z_0)^{2k+m}}\\ 
\nonumber &=& z_0 \sum_{m = 0}^{3k+2n-1}  \frac{\left(\sum_{j=0}^n P_j^{(n,k)}(\nu) \sum_{r=0}^j (-1)^{j-r}\right.
\left(
\begin{array}{c}
 j  \\
 r  
\end{array}
\right)
\left.a_{m-r}^{(k,j)}(\nu)\right)}{(\nu - (\nu-1)z_0)^{2k+m}}\\
 \label{final-exp}
 &=& z_0 \sum_{m = 0}^{3k+2n-1}  \frac{\left(\sum_{j=0}^n P_j^{(n,k)}(\nu) \sum_{r=0}^m (-1)^{j-r}\right.
\left(
\begin{array}{c}
 j  \\
 r  
\end{array}
\right)
\left.a_{m-r}^{(k,j)}(\nu)\right)}{(\nu - (\nu-1)z_0)^{2k+m}}.
\end{eqnarray}
In the second line we applied (\ref{a-vanish}) again to extend the upper bound of the innermost summation on the first line in order to be able to pull this summation to the outside. In the last line, the upper bound of the innermost summation was changed from $j$ in the previous line to $m$. To justify this note that if $m < j$, then the terms with index greater than $m$ in the original sum vanish anyway by (\ref{a-vanish}); on the other hand, if $m>j$ then terms greater than $j$ in the new sum will vanish since
$\left(\begin{array}{c}
 j  \\
 r  
\end{array}\right)=0$ 
if $r>j$ per the usual conventions for binomial coefficients.
\medskip

The proof of Proposition \ref{vanres}(iii) thus comes down to establishing the following vanishing lemma.

\begin{lemma} \label{mainlemma}
\begin{eqnarray} \label{vanishing}
\sum_{j=0}^n P_j^{(n,k)} (\nu) \sum_{r=0}^m (-1)^{j-r} 
\left(
\begin{array}{ccc}
 j\\
 r 
\end{array}
\right) a_{m-r}^{(k,j)}(\nu) = 0
\end{eqnarray}
for $m = 0,1, \dots n-1$.
\end{lemma}
\medskip

 {\bf Proof.} Fix $k>0$. The argument proceeds inductively. The base step is for $n=1$ which implies that $m=0$. So the expression to consider is
\begin{eqnarray*}
&&\sum_{j=0}^1 P_j^{(1,k)} (\nu) (-1)^{j}  
\left(
\begin{array}{ccc}
 j\\
 0
\end{array}
\right) a_{0}^{(k,j)}(\nu)\\
&=&  P_0^{(1,k)}(\nu) a_{0}^{(k,0)}(\nu)  - P_1^{(1,k)}(\nu) a_{0}^{(k,1)}(\nu)\\
&=&  \left(P_0^{(1,k)}(\nu)  + P_1^{(1,k)}(\nu) (2k-1)\right) a_{0}^{(k,0)}(\nu) \,\,\, \mbox{by Lemma \ref{L-coeffs}}\\
&=& \left[1-2k + 1\cdot (2k-1)\right] a_{0}^{(k,0)}(\nu) \,\,\, \mbox{by  (\ref{P-rec})}\\
&=& 0.
\end{eqnarray*}
For the induction step, we assume the lemma is true for $n-1$; i.e., 
\begin{eqnarray*}
\sum_{j=0}^{n-1} P_j^{(n-1,k)} (\nu) \sum_{r=0}^m (-1)^{j-r} 
\left(
\begin{array}{ccc}
 j\\
 r 
\end{array}
\right) a_{m-r}^{(k,j)}(\nu) = 0
\end{eqnarray*}
for $m = 0,1, \dots n-2$, and consider the expressions on the left-hand sides of the putative equations in Lemma \ref{mainlemma}:
\begin{eqnarray*}
&&\sum_{j=0}^n (-1)^{j}  P_j^{(n,k)}  \sum_{r=0}^m (-1)^{r} 
\left(
\begin{array}{ccc}
 j\\
 r 
\end{array}
\right) a_{m-r}^{(k,j)}\\
&=& -(2k-2+n) \sum_{j=0}^{n} (-1)^{j}  P_j^{(n-1,k)}  \sum_{r=0}^m (-1)^{r} \left(
\begin{array}{ccc}
 j\\
 r 
\end{array}
\right) a_{m-r}^{(k,j)}\\
&+&  \sum_{j=0}^{n} (-1)^{j} \left[ P_{j-1}^{(n-1,k)}  +  \left\{(\nu-1) j\right\}P_j^{(n-1,k)}\right]\sum_{r=0}^m (-1)^{r} \left(
\begin{array}{ccc}
 j\\
 r 
\end{array}
\right) a_{m-r}^{(k,j)} \,\,\, \mbox{applying}\,\,\, (\ref{P-recI}) \\
&=& -(2k-2+n) \sum_{j=0}^{n-1} (-1)^{j}  P_j^{(n-1,k)}  \sum_{r=0}^m (-1)^{r} \left(
\begin{array}{ccc}
 j\\
 r 
\end{array}
\right) a_{m-r}^{(k,j)}\\
&-&  \sum_{j=0}^{n-1} (-1)^{j} P_{j}^{(n-1,k)} \sum_{r=0}^m (-1)^{r}\left(
\begin{array}{ccc}
 j+1\\
 r 
\end{array}
\right) a_{m-r}^{(k,j+1)} \\
&+& (\nu-1) \sum_{j=0}^{n-1} (-1)^{j}  P_j^{(n-1,k)}\sum_{r=0}^m (-1)^{r} j \left(
\begin{array}{ccc}
 j\\
 r 
\end{array}
\right) a_{m-r}^{(k,j)} \,\,\, \mbox{by}\,\,\, (\ref{P-recIII}) \,\,\, \mbox{and shifting $j$ in the middle sum}\\
&=& -(2k-2+n) \sum_{j=0}^{n-1} (-1)^{j}  P_j^{(n-1,k)}  \sum_{r=0}^m (-1)^{r} \left(
\begin{array}{ccc}
 j\\
 r 
\end{array}
\right) a_{m-r}^{(k,j)}\\
&-&  \sum_{j=0}^{n-1} (-1)^{j} P_{j}^{(n-1,k)} \sum_{r=0}^m (-1)^{r}\left(
\begin{array}{ccc}
 j+1\\
 r 
\end{array}
\right)\left[ \{j\nu - (2k+m-r+ (j-1))\}a_{m-r}^{(k,j)}\right.\\
&+&\left. \nu \{2k+m-r+(j-1)\} a_{m-r-1}^{(k,j)}\right]\\
&+& (\nu-1) \sum_{j=0}^{n-1} (-1)^{j}  P_j^{(n-1,k)}\sum_{r=0}^m (-1)^{r} j \left(
\begin{array}{ccc}
 j\\
 r 
\end{array}
\right) a_{m-r}^{(k,j)} \,\,\, \mbox{by Lemma \ref{L-coeffs}}\\
&=& -(2k-2+n) \sum_{j=0}^{n-1} (-1)^{j}  P_j^{(n-1,k)}  \sum_{r=0}^m (-1)^{r} \left(
\begin{array}{ccc}
 j\\
 r 
\end{array}
\right) a_{m-r}^{(k,j)}\\
&+& (2k-1+m) \sum_{j=0}^{n-1} (-1)^{j}  P_j^{(n-1,k)}  \sum_{r=0}^m (-1)^{r} \left(
\begin{array}{ccc}
 j+1\\
 r 
\end{array}
\right) \left\{a_{m-r}^{(k,j)}  - \nu a_{m-r-1}^{(k,j)} \right\}\\
&+& \sum_{j=0}^{n-1} (-1)^{j}  P_j^{(n-1,k)}  \sum_{r=0}^m (-1)^{r} (j-r) \left(
\begin{array}{ccc}
 j+1\\
 r 
\end{array}
\right) \left\{a_{m-r}^{(k,j)}  - \nu a_{m-r-1}^{(k,j)} \right\}\\
&-& \nu \sum_{j=0}^{n-1} (-1)^{j}  P_j^{(n-1,k)}\sum_{r=0}^m (-1)^{r} j \left(
\begin{array}{ccc}
 j+1\\
 r 
\end{array}
\right)
a_{m-r}^{(k,j)}\\
&+& (\nu-1)  \sum_{j=0}^{n-1} (-1)^{j}  P_j^{(n-1,k)}\sum_{r=0}^m (-1)^{r} j \left(
\begin{array}{ccc}
 j\\
 r 
\end{array}
\right) a_{m-r}^{(k,j)}.
\end{eqnarray*}
We next rearrange the terms in the last expression and make use of {\it Pascal's identity}, 
$$\left(
\begin{array}{ccc}
 j+1\\
 r 
\end{array}
\right) = \left(
\begin{array}{ccc}
 j\\
 r 
\end{array}
\right) + \left(
\begin{array}{ccc}
 j\\
 r-1 
\end{array}
\right)$$
to achieve some reductions. 
\begin{eqnarray*}
&&\sum_{j=0}^n (-1)^{j}  P_j^{(n,k)}  \sum_{r=0}^m (-1)^{r} 
\left(
\begin{array}{ccc}
 j\\
 r 
\end{array}
\right) a_{m-r}^{(k,j)}\\
&=& \left(m-(n-1)\right) \sum_{j=0}^{n-1} (-1)^{j}  P_j^{(n-1,k)}  \sum_{r=0}^m (-1)^{r} \left(
\begin{array}{ccc}
 j\\
 r 
\end{array}
\right) a_{m-r}^{(k,j)}\\
&+& (2k-2+m) \sum_{j=0}^{n-1} (-1)^{j}  P_j^{(n-1,k)}  \sum_{r=0}^m (-1)^{r} \left(
\begin{array}{ccc}
 j\\
 r-1
\end{array}
\right) \left\{a_{m-r}^{(k,j)}  - \nu a_{m-r-1}^{(k,j)} \right\}\\
&+& \sum_{j=0}^{n-1} (-1)^{j}  P_j^{(n-1,k)}  \sum_{r=0}^m (-1)^{r} j \left(
\begin{array}{ccc}
 j\\
 r 
\end{array}
\right) \left\{a_{m-r}^{(k,j)}  - \nu a_{m-r-1}^{(k,j)} \right\}\\
&-& \nu \sum_{j=0}^{n-1} (-1)^{j}  P_j^{(n-1,k)}\sum_{r=0}^m (-1)^{r} j \left(
\begin{array}{ccc}
 j\\
 r-1
\end{array}
\right)
a_{m-r}^{(k,j)}\\
&-&   \sum_{j=0}^{n-1} (-1)^{j}  P_j^{(n-1,k)}\sum_{r=0}^m (-1)^{r} j \left(
\begin{array}{ccc}
 j\\
 r 
\end{array}
\right) a_{m-r}^{(k,j)}\\
&=& \left(m-(n-1)\right) \sum_{j=0}^{n-1} (-1)^{j}  P_j^{(n-1,k)}  \sum_{r=0}^m (-1)^{r} \left(
\begin{array}{ccc}
 j\\
 r 
\end{array}
\right) a_{m-r}^{(k,j)}\\
&+& (2k-2+m) \sum_{j=0}^{n-1} (-1)^{j}  P_j^{(n-1,k)}  \sum_{r=1}^m (-1)^{r} \left(
\begin{array}{ccc}
 j\\
 r-1
\end{array}
\right) \left\{a_{m-r}^{(k,j)}  - \nu a_{m-r-1}^{(k,j)} \right\}\\
&-& \nu \sum_{j=0}^{n-1} (-1)^{j}  P_j^{(n-1,k)}  \sum_{r=0}^{m-1} (-1)^{r} j \left(
\begin{array}{ccc}
 j\\
 r 
\end{array}
\right)   a_{m-r-1}^{(k,j)}\\
&-& \nu \sum_{j=0}^{n-1} (-1)^{j}  P_j^{(n-1,k)}\sum_{r=1}^m (-1)^{r} j \left(
\begin{array}{ccc}
 j\\
 r-1
\end{array}
\right)
a_{m-r}^{(k,j)}\\
&=& \left(m-(n-1)\right) \sum_{j=0}^{n-1} (-1)^{j}  P_j^{(n-1,k)}  \sum_{r=0}^m (-1)^{r} \left(
\begin{array}{ccc}
 j\\
 r 
\end{array} 
\right) a_{m-r}^{(k,j)}\\
&-& (2k-2+m) \sum_{j=0}^{n-1} (-1)^{j}  P_j^{(n-1,k)}  \left\{ \sum_{r=0}^{m-1} (-1)^{r} \left(
\begin{array}{ccc}
 j\\
 r
\end{array}
\right) a_{m-r-1}^{(k,j)}  - \nu \sum_{r=0}^{m-2} (-1)^{r} \left(
\begin{array}{ccc}
 j\\
 r
\end{array}
\right)  a_{m-r-2}^{(k,j)} \right\}.
\end{eqnarray*}
The second line of this final expression vanishes by induction for all $m \leq n-1$ since the upper bounds of the internal summations, $m-1$ or $m-2$, are $\leq n-2$ in this range.
Similarly, the first line vanishes, for $m \leq n-2$, by induction. For $m=n-1$ it also vanishes because the leading coefficient, $m-(n-1)$, vanishes for this value of $m$. 
This completes the induction and the proof of Lemma \ref{mainlemma}. $\Box$

\section{Applications} \label{applications}

\subsection{Integral and Recursion Formulas for Map Counts} \label{counts}
In this section we will give several representations of the count for two-legged maps (\ref{2legs}). The most direct one will involve the Laurent expansions of $z_g^{(j)}$ that were developed in Section \ref{proofs}.
\smallskip

 We make use of the defining implicit relation, (\ref{II.005}),  for $z_0$ given by
\begin{equation} \label{6_0}
1 = z(s) - \alpha z(s)^\nu
\end{equation}
where $\alpha = c_\nu s$ and 
$$
c_\nu = 2\nu \left(\begin{array}{c}
  2\nu-1\\
  \nu-1\\
\end{array}\right).
$$

The $j^{th}$ coefficient of the Taylor expansion of $z_g$ as a
function of $\alpha$ near $0$, $z_g = \sum_{j\geq 0} \zeta_j^{(g)}(\nu)
\alpha^j$, is naturally given by

\begin{eqnarray} \label{residue}
\zeta_j^{(g)}(\nu) &=& \frac{1}{2\pi i} \oint \frac{z_g(\alpha)}{\alpha^{j+1}}
d\alpha.
\end{eqnarray}

One can change variables from $\alpha$ near $0$, in this integral, to $z$ near $1$ by differentiating the relation (\ref{6_0}) 
\begin{eqnarray*}
\frac{dz}{d\alpha} &=& \frac{z^{\nu}}{1-\nu\alpha z^{\nu -1}}
\end{eqnarray*}
and then using the relation again to eliminate $\alpha$ in the differential
\begin{eqnarray*}
\frac{d\alpha}{dz} &=&  \frac{1-\nu\alpha z^{\nu-1}}{z^\nu} = \frac{\nu - (\nu-1)z}{z^{\nu+1}},
\end{eqnarray*}
so that
\begin{eqnarray*}
\zeta_j^{(g)}(\nu) &=& \frac{1}{2\pi i} \oint_{z\sim1} \frac{(\nu - (\nu-1)z)z^{\nu j-1}z_g(z)}{(z-1)^{j+1}} dz\\
&=& \frac{1}{2\pi i} \oint_{z\sim1} \frac{z^{\nu j}P_{3g-2}(z)}{(z-1)^{j}(\nu - (\nu-1)z)^{5g-2}} dz.
\end{eqnarray*}
In the second line we have rewritten the integrand using the more explicit form of $z_g(z)$ given by Theorem \ref{result}. An alternative expression can be found by using instead the partial fractions expansion of $z_g$ given by (\ref{pole-exp}). 
\begin{eqnarray*}
\zeta_j^{(g)}(\nu) &=& \sum_{i=0}^{3g-1} \frac{1}{2\pi i} \oint_{z\sim1} \frac{a_i^{(g,0)}(\nu) z^{\nu j}}{(y-1)^{j+1}(\nu-(\nu-1)z)^{2g+i-1}} dz
\end{eqnarray*}
where $\sum_{i=0}^{3g-1} a_i^{(g,0)}(\nu)  = 0$ since $z_g(z)$ vanishes at $z=1$ for $g>0$. 
\bigskip

These integral formulas for the generating function coefficients may be re-expressed as recursion formulas. To accomplish this transformation we return to the $\alpha$ contour integral (\ref{residue}) which may be recast in terms of higher derivatives as
\begin{eqnarray} \label{residue2}
\nonumber \zeta_j^{(g)}(\nu) &=& \frac{1}{2\pi i} \frac{1}{j! c_\nu^j} \oint_{s\sim 0} \frac{z_g^{(j)}(s)}{s}ds\\
&=& \frac{1}{2\pi i} \frac{1}{j! c_\nu^j} \oint_{z\sim 1} \frac{(\nu - (\nu-1)z) z_g^{(j)}(z)}{z(z-1)}dz\\
&=& \frac{1}{2\pi i} \frac{1}{j!} \oint_{z\sim 1} \frac{z^{j\nu}}{z-1} \left(\sum_{\ell=0}^{3g-1+ j}\frac{a_\ell^{(g,j)}(\nu)}{(\nu - (\nu-1)z)^{2g-1+j+\ell}}\right) dz.
\end{eqnarray}
where in the second line we applied the Laurent expansion (\ref{zkder-pole}).
\begin{cor}
\begin{eqnarray*}
z_g^{(j)}(0) &=& ^\# \{\mbox{two-legged
$g$-maps with $j$ $2\nu$-valent vertices }\} =  j! c_\nu^{j} \zeta_j^{(g)}(\nu) \\   
&=& c_\nu^{j} \sum_{\ell=0}^{3g-1+ j} a_\ell^{(g,j)}(\nu)\,\,\, \mbox{and}\\
a_\ell^{(g,j)}(\nu) &=&  [(j-1)\nu - (2g + \ell + (j-2))] a_\ell^{(g, j-1)}(\nu) + \nu [2g + \ell + (j-2)] a_{\ell - 1}^{(g, j-1)}(\nu)
\end{eqnarray*}
by Lemma \ref{L-coeffs}  with 
\begin{eqnarray*}
a_0^{(0,0)}(\nu) &=&1\\
a_\ell^{(0,0)}(\nu) &=& 0 \,\,\,\,\, \mbox{for} \,\,\,\,\, \ell > 0\\
a_\ell^{(g,j)}(\nu) &=& 0  \,\,\,\,\, \mbox{for} \,\,\,\,\, \ell < 0
\end{eqnarray*}
and
\begin{eqnarray*}
&& \sum_{\ell=0}^{3g-1} a_\ell^{(g,0)}(\nu)  = 0.
\end{eqnarray*}
The coefficients $a_\ell^{(g,j)}(\nu)$ appearing in this evaluation are thus recursively expressible in term of the $3g$ coefficients, $a_\ell^{(g,0)}(\nu)$, of the partial fractions expansion of $z_g$. These latter coefficients are rational functions of the 
$d_\lambda^{(\nu,k)}$, for $k \leq g$,
over the rational numbers $\mathbb{Q}$.
\end{cor}

\subsection{Rationality of the Asymptotic Partition Function Coefficients} \label{partfcn-asymp}

We will now make use of the following recursive integral formulae for the fundamental coefficients $e_g$. (The derivation of these formulae is accomplished by a process similar to that described in section \ref{continuum} for the coefficients $z_g$; for the details we refer the reader to \cite{EMP}.) 

\begin{thm}\label{II.004thm}  \cite{EMP}

\begin{align} \label{egform}
e_g(-s) &=
 - \frac{1}{(2-2g)(1-2g)}
  \mbox{\textup{drivers}}_g(z_0(s))
 \\ \nonumber &  - \frac{1}{2-2g} \left( \frac{z_0(s)-1 }{c_\nu z_0(s)^\nu} \right)^{(2g-2)/(\nu-1)} \int_1^{z_0(s)}
\left( \frac{c_\nu z^\nu}{z-1} \right)^{(2g-2)/(\nu-1)} \left(
\mbox{\textup{drivers}}_g(z) \right)^{\bullet} dz  \\
\nonumber&  + \frac{1}{(1-2g)} \left( \frac{z_0(s)-1}{c_\nu z_0(s)^\nu} \right)^{(2g-1)/(\nu-1)} \int_1^{z_0(s)}
\left( \frac{c_\nu z^\nu}{z-1} \right)^{(2g-1)/(\nu-1)}
\left(\mbox{\textup{drivers}}_g(z) \right)^{\bullet} dz
\\ \nonumber &
+ K_1 s^{(2g-2)/(\nu-1)} + K_2 s^{(2g-1)/(\nu-1)}
\end{align}
when $g \neq 1$, where 
\begin{eqnarray}\label{II.007}
\mbox{\textup{drivers}}_g(z_0(s)) &=& -\sum_{\ell
=1}^{g} \frac{2}{(2\ell+2)!} \frac{\partial^{(2\ell+2)}}{\partial
  w^{(2\ell+2)}}
\left[
  w^{2-2(g-\ell)} e_{g-\ell}(-w^{\nu-1} s)  \right] \bigg|_{w=1}\\
  \nonumber  &+& \mbox{the}\;
  n^{-2g} \; \mbox{term of} \; \log\left( \sum_{m = 0}^\infty \frac{1}{n^{2m}}
  z_m(s) \right).
\end{eqnarray} 
We denote by $\left( \mbox{\textup{drivers}}_g(z)\right)^{\bullet}$ the derivative of $ \mbox{\textup{drivers}}_g(z)$ with respect to $z$.
$K_1$ and $K_2$ are constants of
integration either determined by the requirement that $e_g$ be a
locally analytic function of $s$ or by the evaluation of $e_g$ for
low values of $\nu$ through its combinatorial characterization.
When $g=1$,
\begin{align} \nonumber
e_1(-s) &= \frac{1}{(\nu-1)} \left[ \left( \frac{z_0(s)-1}{c_\nu
    z_0(s)^\nu} \right)^{1/(\nu-1)} \int_1^{z_0(s)} \left( \frac{c_\nu
    z^\nu}{z-1} \right)^{\nu/(\nu-1)} \frac{(\nu - (\nu-1) z)}{c_\nu
    z^{\nu+1} } \mbox{\textup{drivers}}_1(z) dz
\right. \\ \nonumber &\phantom{= \frac{1}{(\nu-1)} } \left.
-\int_1^{z_0(s)} \frac{ (\nu - (\nu-1) z)}{z (z-1)}
    \mbox{\textup{drivers}}_1(z) dz \right] \\
\label{e1form}
&= -\frac{1}{12} \log\left( \nu - (\nu-1) z_0(s) \right),
\end{align}
where we have chosen the principal branch of the logarithm.

For convenience we also record here the planar ($g=0$) result,
\begin{equation} \label{e0form}
e_0(-s) =  \frac{1}{2} \log(z_0(s)) + \frac{(\nu-1)^2}{4\nu(\nu+1)} (z_0(s) - 1)
\left(z_0(s) - \frac{3(\nu+1)}{\nu-1} \right).
\end{equation}
\end{thm}
\medskip

 We can now extend this characterization by proving Theorem \ref{egratl}. 
\bigskip

 {\it For $g \geq 2$, the coefficient $e_g(-s) = e_g(z_0(s))$ is a rational function of $z_0$  with poles only at $z_0 =\nu/(\nu - 1)$ and which vanishes at least simply at $z_0 = 1$.}
\medskip

 {\bf Proof.} We proceed inductively assuming the theorem to hold for all $e_r$ with $2 \leq r < g$ and with $e_1$ and $e_0$ as given, respectively, by (\ref{e1form}) and (\ref{e0form}). To start, one needs to examine the terms of 
$\mbox{\textup{drivers}}_g(z)$ for $g \geq 2$. Note that for $1 \leq \ell \leq g-2$,
\begin{eqnarray} \label{wder}
&&\frac{\partial}{\partial w} \left[w^{2-2(g-\ell)} e_{g-\ell}(-w^{\nu-1} s)  \right]\\
\nonumber &=& \left(2-2(g-\ell)\right) w^{1-2(g-\ell)} e_{g-\ell} \left(-w^{\nu-1}s\right) + w^{\nu -2(g-\ell)} s e_{g-\ell}^\bullet \left(-w^{\nu-1}s\right)\\
\nonumber &=& \left(2-2(g-\ell)\right) w^{1-2(g-\ell)} e_{g-\ell} \left(-w^{\nu-1}s\right) + w^{\nu -2(g-\ell)} \frac{z_0-1}{c_\nu z_0^\nu} e_{g-\ell}^\bullet \left(-w^{\nu-1}s\right)
\end{eqnarray}
where in the third line we have used the identity (\ref{ess}).
The final expression for (\ref{wder}) has three properties that will be useful to note:
\begin{eqnarray} 
&&\label{properties1} \mbox{It vanishes at} \,\,\,  z_0 = 1.\\
&&\label{properties2} \mbox{Its minimal pole order at} \,\,\,  z_0 = \nu/{\nu-1} \,\,\,  \mbox{is not less than that of} \,\,\, e_{g-\ell}(z).\\
&& \label{properties3} \mbox{It may also have poles at} \,\,\,  z_0 = 0 \,\,\, \mbox{but nowhere else}.
\end{eqnarray}
It is straightforward to see that these three properties are maintained under further differentiation with respect to $w$. 

We next separately check that these same properties hold for $w$-derivatives of $e_{g-\ell}$ when $\ell = g-1$ or $g$, which correspond respectively to $e_1$ and $e_0$. For $e_1$ we have
\begin{eqnarray*}
\frac{\partial}{\partial w} \left[e_{1}(-w^{\nu-1} s)  \right]
&=& \frac{\nu-1}{12} \cdot \frac{w^{\nu-2}}{\nu - (\nu-1)z_0}\cdot \frac{z_0-1}{c_\nu z_0^\nu}.
\end{eqnarray*}
It is clear from this that the $2g^{th}$ $w$-derivative of $e_1\left(-w^{\nu-1} s\right)$ appearing in $\mbox{\textup{drivers}}_g$ has the three properties (\ref{properties1} - \ref{properties3}). Similarly one can see from (\ref{e0form}) that the $(2g+2)^{th}$ $w$-derivative of $w^2 e_0\left(-w^{\nu-1} s\right)$ appearing in $\mbox{\textup{drivers}}_g$ has the three properties (\ref{properties1} - \ref{properties3}). Property (\ref{properties2}) is vacuous in this case since the expression has no poles at $z_0 = \nu/{\nu-1}$; we take the minimal pole order in this case to be $\infty$.
\smallskip

Lastly we consider the last collection of terms in $\mbox{\textup{drivers}}_g$.
\begin{eqnarray*}
\log\left( \sum_{m = 0}^\infty \frac{1}{n^{2m}}z_m(s) \right) &=& \log(z_0(s)) + \log\left( 1 + \sum_{m = 1}^\infty \frac{1}{n^{2m}}\frac{z_m(s)}{z_0(s)} \right). 
\end{eqnarray*}
The $n^{-2g}$ term of this expression has terms of the form, for $m_1 + \dots + m_r = g$,
\begin{eqnarray*}
\frac{z_{m_1} \dots z_{m_r}}{z_0^r} &=& \frac{\left(z_0 - 1\right)^r Q_{3g-2r}(z_0)}{\left(\nu - (\nu - 1)z_0\right)^{5g-r}}
\end{eqnarray*}
where $Q_{3g-2r}$ is a polynomial of degree $3g-2r$. The form of the right hand side follows directly from Theorem \ref{result} and shows that this part of $\mbox{\textup{drivers}}_g$ vanishes at $z_0 =1$, has minimum pole order of $2g$ at 
$z_0 = \frac{\nu}{\nu-1}$ and of order $\infty$ at $z_0 = 0$.  We may thus conclude that 
\begin{lemma}
For $g \geq 2$, $\mbox{\textup{drivers}}_g(z)$ vanishes at least simply at $z_0 =1$, and its only singularities are poles restricted to occur at either $z =0$ or $z = \nu/(\nu-1)$. 
\smallskip 

 Also, $\mbox{\textup{drivers}}_g(z)^\bullet $ has only poles for singularities which are restricted to occur at either $z =0$ or $z = \nu/(\nu-1)$ with minimal pole order at least 2 at both locations.
 \end{lemma}
We also observe, as was noted in \cite{EMP}, that the terms of (\ref{egform}) involving $K_1$ and $K_2$ can only have the possibility to contribute if $(2g-2)/(\nu - 1)$ or $(2g-1)/(\nu-1)$ are integers.   But in that case it follows from (\ref{ess}) that they are terms which vanish simply at $z_0 =1$ and have poles only at $z_0 = 0$. As a consequence of this observation and the lemma one may conclude from the representation (\ref{egform}) that $e_g$, for $g \geq 2$, is locally rational at $z_0 = \frac{\nu}{\nu-1}$. It thus suffices to consider the behavior of  (\ref{egform}) near $z_0 = 0,1$ in order to establish Theorem \ref{egratl}.

Returning to Theorem \ref{EQMSTHM}, note that when ${t} = 0, \log \left(\frac{Z^{(k)}_{N}({t})}{Z^{(k)}_{N}(\bf{0})} \right) = 0$; therefore, $e_g = 0$ when $s=0$ (or equivalently when $z_0 =1$) for all $g$. For the integral terms in (\ref{egform}), since by the preceding lemma $\mbox{\textup{drivers}}_g(z)^\bullet $ is regular at $z=1$, one may write the integrand in each of these terms as a Puiseux series around $z=1$.  The integral can then be carried out term by term on this series with constant of integration set to zero (note that the constants of integration are already captured in the terms containing the constants $K_1$ and $K_2$ . However, the prefactor in front of each of these integrals reduces the total result to a Laurent series at $z_0 = 1$ except for the possibility of one term proportional to $(z_0-1)^m\log(z_0-1)$. However, since by Theorem \ref{EQMSTHM} $e_g$ is analytic near $z_0 =1$, such a term must either cancel with other such terms in the full expression for $e_g$, or it cannot exist. It follows that (\ref{egform}) has a regular Laurent expansion at $z_0 = 1$ which in fact must be a Taylor expansion with vanishing constant term since $e_g = 0$ at $z_0 = 1$.  

We apply Theorem \ref{extension} once more to conclude that $e_g$ is regular at $z_0 = 0$:
\begin{eqnarray*}
e_g(t) &=& N^{2g}\left[N^{-2} \log \left(\frac{Z_{N}(t)}{Z_{N}(0)} \right) -  e_0(t) - \dots - N^{-(2g-2)}e_{g-1}(t)\right]\\
&-&  N^{2g}\left[N^{-2} \log \left(\frac{Z_{N}(t)}{Z_{N}(0)} \right) - e_0(t) - \dots - N^{-2g}e_{g}(t)\right] \\
\implies  |e_g(t)| &\leq& C_{g-1} + \frac{C_{g}}{N^{2}}
\end{eqnarray*}
uniformly as $t \to \infty$ and hence $e_g$ remains bounded as $z_0$ approaches $0$. It follows, in exactly the same manner as was done for $z_0$ near $1$, that all singular parts of (\ref{egform}) at $z_0 = 0$ must cancel each other. $\Box$
\begin{cor}
The integration constants $K_1$ and $K_2$ can be non-zero only if $(2g-2)/(\nu - 1)$ or $(2g-1)/(\nu-1)$ are integers and in addition the combination of the {\it other} terms in (\ref{egform}) has a pole at $z_0 = 0$.
\end{cor}

\subsection{The Double Scaling Limit for $\nu = 2$ and Painlev{\'e} I } \label{dsI}

The topic of the so-called {\it double-scaling limit} of the random matrix partition function (\ref{I.001}) has been extensively discussed in both the physics and the mathematical physics literature related to two-dimensional quantum gravity. For a concise history of these developments and relevant references we refer the reader to \cite{FIKN}.  In this literature attention was focused on the special case of the weight $V(\lambda) =  \frac{1}{2} \lambda^{2} + t \lambda^{4}$ which corresponds to the case of $\nu = 2$ in our more general treatment.  In this case the double-scaling limit refers to the regime in parameter space where $t \to t_c (=\frac{-1}{48})$ and $N \to \infty$ simultaneously so that
\begin{eqnarray} \label{ds}
N^{4/5} \left( t + \frac{1}{48}\right) &=& - \gamma_1 \,\, \xi
\end{eqnarray}
remains fixed. The authors in \cite{BreKaz, DoS, GrMi} arrived at this prescription of the limit through a formal scaling argument based on the Freud equation (\ref{string}). Basically they posited a general scaling form, in $n, N$ and $t$, for the recurrence coefficients $b^2_{n,N}(t)$ and determined that (\ref{ds}) was the unique scaling that would (formally) asymptotically balance (\ref{string}).  Realizing that this relation, which they called the string equation, could be taken to be a discrete form of the first Painlev{\'e} equation
\begin{eqnarray} \label{PI}
y^{\prime\prime} = 6 y^2 + \xi,
\end{eqnarray}
they took (\ref{PI}) to represent the exact {\it non-perturbative string theory} for two-dimensional euclidean quantum gravity. 

The corresponding mathematical conjecture was that the large $N$ expansion of the recurrence coefficients, under the scaling (\ref{ds}), has a limiting form $y(\xi)$, at leading order in $N$, that solves (\ref{PI}) which hereafter will be referred to as PI. The proposition was first placed on a rigorous footing, based on Riemann-Hilbert analysis, in the papers \cite{FIKI, FIKII}. A complete and self-contained proof was recently detailed in \cite{DK}.
\medskip

The Freud equation and its relation to PI play no role in our description of the double scaling limit for general $\nu$ which we take up in the next subsection. The reason that we have dwelled on them for the case $\nu = 2$ in this subsection, apart from their historical relevance for our topic, is that this special case provides a concrete point of comparison (see Section \ref{dsII}) for our more general results {\it and} a jumping off point for possible future investigations that might involve a broader connection to Painlev{\'e} transcendents.  For these reasons we will briefly review, here,  the essential properties of the relevant PI solutions. For a systematic development and further details we refer the reader to \cite{Kap}. 

There is a family of solutions to (\ref{PI}) characterized by having the common asymptotic expansion
\begin{eqnarray} \label{PI-exp}
y(\xi) &\sim& \sqrt{\frac{-\xi}{6}} \left( 1 + \sum_{g=1}^\infty \alpha_g (-\xi)^{-5g/2}  \right) \,\,\, \mbox{as} \,\,\, \xi \to -\infty
\end{eqnarray}
for particular coefficients $\alpha_g$. These expansions are valid in angular sectors, typically of width $\frac{2\pi}{5}$ in the complex $\xi$ plane. They differ from one another by jumps across sectors (Stokes lines) which are exponentially small. When needed, for specificity, we will take $y(\xi)$ to be that solution with the above asymptotic expansion valid in the full sector $\frac{3\pi}{5} < \arg \xi < \frac{7\pi}{5}$ as $|\xi| \to \infty$. 
The $\alpha_g$ satisfy the following quadratic recursion formula
\begin{eqnarray} \label{PI-rec}
\alpha_{g+1} &=& \frac{25 g^2 - 1}{8\sqrt{6}} \alpha_g - \frac{1}{2} \sum_{m=1}^g \alpha_m \alpha_{g+1-m}\\
\nonumber \alpha_0 &=& 1.
\end{eqnarray}

\subsection{The Double Scaling Limit for General $\nu$} \label{DS}
In this section we will describe a double scaling limit for the free energy of the unitary ensembles (\ref{RMT}) in terms of the rational representation of the $z_g$. This will be done in the setting of general values of $\nu$ which should enable one to see the universal character of these results.   

We return to the asymptotic expansion for the orthogonal polynomial recursion coefficients (Theorem \ref{II.002} or Theorem \ref{extension}) and substitute the polar expansions found in Theorem \ref{result}.
\begin{eqnarray}
&&b_{N,N}^2 = z_0 + \sum_{g=1}^\infty z_g N^{-2g}\\
\nonumber && b_{N,N}^2 - \frac{\nu}{\nu-1} = \left( z_0 - \frac{\nu}{\nu-1}\right) + \sum_{g=1}^\infty z_g N^{-2g}\\
\label{rec-exp}&& = \left( z_0 - \frac{\nu}{\nu-1}\right)\\
\nonumber && +  z_0 \sum_{g=1}^\infty \left\{ \frac{a_0^{(g)}(\nu)}{(\nu - (\nu-1)z_0)^{2g}} + \frac{a_1^{(g)}(\nu)}{(\nu - (\nu-1)z_0)^{2g+1}}+ \cdots + \frac{a_{3g-1}^{(g)}(\nu)}{(\nu - (\nu-1)z_0)^{5g-1}}\right\} N^{-2g}.
\end{eqnarray}
We first proceed formally to determine what the double scaling balance should be; i.e., we choose $\delta$ in
\begin{eqnarray*}
\nu - (\nu-1) z_0 &\sim& N^\delta
\end{eqnarray*}
such that the highest order terms of (\ref{rec-exp}) at the pole have a common factor involving $N$ that is independent of $g$. In other words 
$\delta$ should be chosen so that $\lfloor(5g-1)\delta\rfloor = -2g$. Therefore, $\delta = -\frac{2}{5}$. Combining this with (\ref{caustic2}) one is led to take 
the double scaling ansatz for general $\nu$ to be
\begin{eqnarray} \label{ds-nu}
N^{4/5} \left( s - s_c \right) &=& \gamma^{(\nu)}_1 \,\, \xi
\end{eqnarray}
where, by (\ref{caustic}), $s_c = \frac{(\nu -1)^{\nu-1}}{c_\nu \nu^\nu}$ and $\gamma^{(\nu)}_1$ is a constant depending only on $\nu$. We will in general notationally suppress the explicit dependence of $\gamma_1$ on $\nu$ since this should be clear from context.  Also, by (\ref{caustic2}), 
\begin{eqnarray*}
\left( z_0 - \frac{\nu}{\nu-1}\right) &=& - \sqrt{\frac{-2 c_\nu \nu^{\nu+1}}{(\nu - 1)^\nu}(s - s_c)} \left\{1 + \mathcal{O}(s - s_c)\right\}\\
&=& - \sqrt{\frac{-2 c_\nu \nu^{\nu+1}}{(\nu - 1)^\nu}\gamma_1 \, \xi} \,\, N^{-2/5} + \mathcal{O}(N^{-4/5}) \,\,\, \mbox{by} \,\,\, (\ref{ds-nu}).
\end{eqnarray*}
Substituting the ansatz (\ref{ds-nu}) into (\ref{rec-exp}), one may formally conclude that 
\begin{eqnarray} \label{rec-exp-final}
\nonumber && b_{N,N}^2 - \frac{\nu}{\nu-1} \\
\nonumber && = - \sqrt{\frac{-2 c_\nu \nu^{\nu+1}}{(\nu - 1)^\nu}\gamma_1 \xi} \,\, N^{-2/5}\\
\nonumber &&+  \frac{\nu}{\nu-1} \sum_{g=1}^\infty \left\{ \frac{a_0^{(g)}/(\nu-1)^{2g}}{\left( \frac{-2 c_\nu \nu^{\nu+1}}{(\nu - 1)^\nu}\gamma_1 \xi \right)^{g}N^{-4g/5}} +  \cdots + \frac{a_{3g-1}^{(g)}/(\nu-1)^{5g-1}}{\left(\frac{-2 c_\nu \nu^{\nu+1}}{(\nu - 1)^\nu}\gamma_1 \xi \right)^{(5g-1)/2}N^{-2g +2/5}}\right\} N^{-2g}\\
\nonumber &&+ \mathcal{O}(N^{-4/5})\\
\nonumber && = - \sqrt{\frac{-2 c_\nu \nu^{\nu+1}}{(\nu - 1)^\nu}\gamma_1 \xi} \,\, N^{-2/5}
+  \frac{\nu}{\nu-1} \sum_{g=1}^\infty \frac{a_{3g-1}^{(g)}}{\left(\frac{-2 c_\nu \nu^{\nu+1}}{(\nu - 1)^{\nu+2}}\gamma_1 \xi \right)^{(5g-1)/2}} N^{-2/5}\\
\nonumber &&+ \mathcal{O}(N^{-4/5})\\
\nonumber && = - \sqrt{\frac{-2 c_\nu \nu^{\nu+1}}{(\nu - 1)^\nu}\gamma_1 \xi} \,\, N^{-2/5}
+  \frac{\nu}{\nu-1} \sum_{g=1}^\infty \frac{a_{3g-1}^{(g)}}{\left(\frac{-2 c_\nu \nu^{\nu+1}}{(\nu - 1)^{\nu+2}}\gamma_1 \xi\right)^{(5g-1)/2}}  N^{-2/5}\\
\nonumber &&+ \mathcal{O}(N^{-4/5})\\
&& = - \left(\frac{-2 c_\nu \gamma_1\nu^{\nu+1}}{(\nu - 1)^{\nu+2}} \xi \right)^{1/2} \left\{
 (\nu - 1) -  \frac{\nu}{\nu-1} \sum_{g=1}^\infty a_{3g-1}^{(g)}(\nu)\left(\frac{-2 c_\nu \gamma_1 \nu^{\nu+1}}{(\nu - 1)^{\nu+2}} \xi\right)^{-5g/2} \right\} N^{-2/5}\\
 \nonumber && + \mathcal{O}(N^{-4/5}).
\end{eqnarray}

We next study what can be said about the leading order, for large $N$, term in the formal expansion (\ref{rec-exp-final}). The coefficients of the series at this leading order  are essentially given by the $a_{3g-1}^{(g)}(\nu)$
which are, sequentially, the highest order coefficients of the pole at $\nu/(\nu - 1)$ for the sequence of $z_g$. We derive a recursion formula for these coefficients. 
\begin{prop} \label{rec-form}
\begin{eqnarray} \label{rec-form(a)}
a^{g+1}_{3(g+1)-1} (\nu) &=& \frac{\nu^3 \left(25g^2 - 1\right)}{6} a_{3g-1}^{(g)}(\nu) + \frac{\nu}{2} \sum_{m=1}^g  a_{3m-1}^{(m)}(\nu) a_{3(g-m+1)-1}^{(g-m+1)}(\nu)\\  \label{rec-form(b)}
a_{2}^{(1)}(\nu) &=& \frac{\nu^2}{6}
\end{eqnarray}
\end{prop}

{\bf Proof.}  First recall from the proof of Theorem \ref{result}, given in Section \ref{ratl-rec} that the coefficient, that $a_{3g-1}^{(g)}(\nu)$, has (inductively) two sources. The first of these comes from forcing the integral equation for $z_g$ (see Theorem \ref{emp}:(iii)) with terms of the form
\begin{equation*} 
\left[\nu c_\nu f_0^{\nu-1} \sum_{m=1}^{g-1}  f_{m}\left( f_{g-m}\right)_w\right]_{w=1}
\end{equation*}
(the $J=1$ case). The other comes from forcing with terms of the form
\begin{equation*} 
\left[\frac{\nu c_\nu}{6} f_0^\nu \left(f_{g-1}\right)_{www}\right]_{w=1}
\end{equation*}
(the $J=3$ case; note that here we have used Proposition \ref{d-coeffs} to calculate that $d^{(\nu,g)}_{(3)} = \nu c_\nu$).

Inserting the partial fractions expansion of these terms as the forcing in the integral equation for $z_g$ and then collecting the coefficients of the 
terms proportional to $1/(\nu - (\nu-1)z_0)^{5g-1}$ one derives an equation involving a mixture of coefficients of the form $a^{(k,j)}_{\ell} (\nu)$.
However, by repeated application of the linear recursions from Lemma \ref{L-coeffs}, supplemented by (\ref{P-recI}) - (\ref{P-recIII}),  one arrives at the following expression involving only the maximal pole coefficients of undifferentiated $z_n$,
\begin{eqnarray*}
\frac{\nu}{\nu-1} a^{g}_{3g-1} (\nu) &=& \frac{\nu^2}{2(\nu-1)} \sum_{m=1}^{g-1} a^{m}_{3m-1} (\nu) a^{g-m}_{3(g-m) - 1} (\nu) \\
                                                              &+&  \frac{\nu^4 (5g-4)(5g-6)}{6(\nu-1)} a^{g-1}_{3(g-1)-1} (\nu).
\end{eqnarray*}
Making some obvious cancellations and shifting from $g$ to $g+1$ one thus deduces (\ref{rec-form(a)}). The initial condition (\ref{rec-form(b)}) follows directly from examination of (\ref{z1}). $\Box$ 

Using (\ref{rec-form(b)}) one can rewrite (\ref{rec-form(a)}) as 
\begin{eqnarray*} 
a^{g+1}_{3(g+1)-1} (\nu) &=& \frac{\nu^3 \left(25g^2\right)}{6} a_{3g-1}^{(g)}(\nu) + \frac{\nu}{2} \sum_{m=2}^{g-1}  a_{3m-1}^{(m)}(\nu) a_{3(g-m+1)-1}^{(g-m+1)}(\nu)
\end{eqnarray*}
It is immediate from this that $a_{3g-1}^{(g)}(\nu) > 0$ and hence,
\begin{cor}
The pole order of $z_g$ at $z_0 = \nu/(\nu - 1)$ is exactly equal to $5g-1$ for all $g \geq 1$.
\end{cor}
It also follows from this form of the recursion that 
$$
a^{g+1}_{3(g+1)-1} (\nu)/ a_{3g-1}^{(g)}(\nu) > C g^2
$$
and therefore the series in the leading order term of (\ref{rec-exp-final}) is divergent as $\xi \to -\infty$.  One expect this series to be an asymptotic and their are various approaches one could take to determine to what functions (as $\nu$ varies) it might be asymptotic (generalized Borel summation, seeking an ode with a solution whose asymptotic coefficients satisfy Proposition \ref{rec-form}, etc.).  However, as we will suggest in the next section, the form of our recursion suggests another approach both to characterizing the leading order series and to establishing the validity of the double scaling expansion (\ref{rec-exp-final}).
\subsection{Universality and the Painlev{\'e} I  Hierarchy} \label{dsII}
We will now restrict our study of  (\ref{rec-exp-final}) to the special case $\nu = 2$ where it specializes to the following, still formal, expansion:
\begin{eqnarray} \label{rec-exp-2}
\nonumber &&  b_{N,N}^2 - 2 \\
&& = - \left(-192 \gamma_1 \xi\right)^{1/2} \left\{1 -  2 \sum_{g=1}^\infty a_{3g-1}^{(g)}\left(-192 \gamma_1  \xi \right)^{-5g/2} \right\} N^{-2/5} + \mathcal{O}(N^{-4/5}).
\end{eqnarray}
We quote a recent result of Duits and Kuijlaars which will enable us to give a precise characterization of (\ref{rec-exp-2}). 
\begin{thm} \cite{DK} \label{dk}
There are constants $\gamma_1$ --see (\ref{ds})-- and $\gamma_2$ such that  
\begin{eqnarray} \label{DK}
b_{N,N}^2 - 2 &=& - \gamma_2 \left( y_\alpha(\xi) + y_\beta(\xi) \right) N^{-2/5} + \mathcal{O}(N^{-3/5}) \,\,\, \mbox{as} \,\,\, N  \to \infty,
\end{eqnarray}
and where $y_\alpha$ and $y_\beta$ are two members of the family of PI solutions having the common asymptotics (\ref{PI-exp}). The constants 
$\gamma_1$ and $\gamma_2$ are independent of the choice of $\alpha$ and $\beta$.
This expansion holds uniformly for $\xi$ in compact subsets of $\mathbb{R}$ not containing any of the poles of  $y_\alpha$ and $y_\beta$ and can in fact be extended to a full asymptotic expansion in powers $N^{-1/5}$.
\end{thm}

The derivation of the above result depends fundamentally on an extension of the definition of orthogonal polynomials with exponential weights to a more general class of non-Hermitean orthogonal polynomials which corresponds, in an appropriate sense, to taking $t$ in the weight $V$ to lie in the interval $\left(\frac{-1}{48}, 0\right)$. There is a corresponding extended notion of the equilibrium measure mentoned in Section \ref{z0-eqmeas} for these negative values of $t$. In \cite{DK} this measure is defined through a variational problem along a deformed contour in the complex $\lambda$ plane. These orthogonal polynomials are characterized by a Riemann-Hilbert (R-H) problem and their asymptotics may be studied via the Deift-Zhou method of nonlinear steepest descent \cite{DKMVZ}. In particular, this involves the construction of local parametrices in terms of the Riemann-Hilbert problem for the PI equation.  (The parameters $\alpha$ and $\beta$ in the above Theorem extend through this process from the contour deformations to the R-H problem for the orthogonal polynomials to the R-H problem for PI to the family of PI solutions. A key point for relating this analysis to our analysis in the case of (\ref{rec-exp-2}) is that for $\frac{-1}{48} < t < 0$ there is an $N_0(t)$ such that for all $N > N_0(t)$ the recurrence coefficients $\hat{b}^2_{N,N}(t)$ for the non-Hermitean polynomials exist (\cite{DK}, Theorem 1.1).  Moreover,  $\hat{b}^2_{N,N}(t)$ has a full asymptotic expansion, independent of the choice of $\alpha$ and $\beta$, in inverse powers of $N$. Indeed the construction of this expansion is modeled on that of \cite{EM}. It follows that $\hat{b}^2_{N,N}(t) = b^2_{N,N}(t)$ for $\frac{-1}{48} < t < 0$.

We can use these observations to directly relate the highest order polar expansion coefficients,  $a_{3g-1}^{(g)}$ of $z_g$ for $\nu = 2$ that appear in (\ref{rec-exp-2}) to the coefficients in the asymptotic expansion (\ref{PI-exp}) of the PI solution. First we need to determine the pinning constants $\gamma_1$ and $\gamma_2$. Matching the coefficients of $N^{-2/5}$  between (\ref{rec-exp-2}), (\ref{DK}) and (\ref{PI-exp}) requires
\begin{eqnarray}
\nonumber - \left(-192 \gamma_1 \xi\right)^{1/2} \left\{1 -  2 \sum_{g=1}^\infty a_{3g-1}^{(g)}\left(-192 \gamma_1  \xi \right)^{-5g/2} \right\} &=& - \gamma_2 \left( y_\alpha(\xi) + y_\beta(\xi) \right)\\
\label{match} &=& -2 \gamma_2 \sqrt{\frac{-\xi}{6}} \left( 1 + \sum_{g=1}^\infty \alpha_g (-\xi)^{-5g/2}  \right).
\end{eqnarray}
Comparison of the first two terms gives
\begin{eqnarray*}
192 \gamma_1 &=& \frac{4}{6} \gamma_2^2\\ 
-\frac{2 a_{2}^{(1)}}{(192 \gamma_1)^{5/2}} &=& \alpha_1\\ 
\alpha_1 &=& -\frac{1}{8\sqrt{6}} \,\,\, \mbox{by} \,\,\, (\ref{PI-rec})\\
 a_{2}^{(1)} &=& \frac{2}{3} \,\,\, \mbox{by} \,\,\, (\ref{rec-form(b)}).
\end{eqnarray*}
From this one immediately sees that 
\begin{eqnarray}
 \label{pin1} \gamma_2 &=& 2^{3/5} 3^{2/5}\\
 \label{pin2} \gamma_1 &=&  \frac{2^{-9/5} 3^{-6/5}}{4} = \frac{1}{4} \gamma_2^{-3}
\end{eqnarray}

These values agree with the pinning constants stated in \cite{DK} (Theorem 1.2) modulo the factor of $1/4$ in (\ref{pin2}). However, in that reference the form of the weight is taken to be $V = \lambda^2/2 + t \lambda^4/4$ which differs from the form of the weight used here effectively by scaling the time $t$ by $1/4$; hence, the first  pinning constant defined in (\ref{ds}) should differ in fact from that appearing in \cite{DK} by a factor of $1/4$. 

With these constants determined one may now use (\ref{match}) to express the highest order pole coefficients (for $\nu = 2$) in terms of the PI asymptotic coefficients. 
\begin{cor}
\begin{eqnarray} \label{hop}
a_{3g-1}^{(g)} &=& - 2^{5g-1} \left( 2/3\right)^{g/2} \alpha_g \,\,\, \mbox{for} \,\,\, g \geq 1.
\end{eqnarray}
\end{cor}
Substitiuting (\ref{hop}) into the nonlinear recusrsion (\ref{PI-rec}) one deduces the following quadratic recursion between the coefficients of the highest order poles of the $z_g$ in the case of $\nu = 2$.
\begin{eqnarray*} 
a_{3(g+1) - 1}^{(g+1)} &=& \frac{4}{3} \left(25 g^2 - 1\right) a_{3g-1}^{(g)} + \sum_{m=1}^g a_{3m-1}^{(m)} a_{3(g+1-m) - 1}^{(g+1-m)}.
\end{eqnarray*}
This is in complete agreement with Proposition \ref{rec-form}  for the case $\nu=2$ and so we can finally state
\begin{prop} \label{validity}
The leading order series in (\ref{rec-exp-2}) coincides with the asymptotic expansion of the PI solution specified in Theorem \ref{dk}. Moreover,
based on the validity of the expansion (\ref{DK}), the expansion (\ref{rec-exp-2}) is also valid for large $N$. 
\end{prop}

In addition, from the derivation of the expansion displayed in (\ref{rec-exp-final}) one deduces an improvement of the result stated in Theorem \ref{dk}.
\begin{cor}
The double-scaling limit of $b_{N,N}^2$ has a full asymptotic expansion in {\it even} powers of $N^{-1/5}$. In particular, the next order correction in (\ref{DK}), after leading order, is $\mathcal{O}(N^{-4/5})$.
\end{cor}

The form of the expansion (\ref{rec-exp-final}) suggests that Proposition \ref{validity}, for the special case of $\nu=2$, should extend to have a universal character for general $\nu$. By {\it universal} here we mean that the various constants and expressions we have been discussing when $\nu =2$  should be replaced in the general case by fixed rational functions of the parameter $\nu$. Although we do not carry out the details here, this extension should follows straightforwardly by generalizing the Riemann-Hilbert problem for PI to the R-H problem for the PI hierarchy.  (For a description of this hierarchy we refer the reader to \cite{Shimomura}.) Specifically one needs to replace the Hamiltonian for PI appearing in the RH problem by the Hamiltonian for the higher order equations in the PI hierarchy. 
\medskip

Another intriguing problem is to study the fine structure of the higher order terms in the double scaling expansion (\ref{DK}); i.e., the coefficients of $N^{-2h/5}$ for integer $h > 1$. This should lead to extensions of Proposition \ref{rec-form} that enable one to recursively determine the fundamental coefficients $a_\ell^{(g)}(\nu)$ for lower values of $\ell$.. 

\subsection{Relation to Other Enumerative Results} \label{other}
The subject of map enumeration has been much studied going back at least to the early work of Tutte \cite{Tu}, in the '60s, who introduced the notion of a rooted map. A map is said to be {\it rooted} if a vertex of the map together with an edge adjacent to it and a side of that edge are distinguished. Since then, much work been done on counting various classes of rooted planar ($g=0$) maps. The cases of higher genus maps have been more challenging and therefore the main emphasis has been placed on trying to determine the asymptotic behavior of various types of enumerations for large values of the discrete parameters.. A particular example of this is given by the problem of counting the number, $M_{n,g}$, of rooted maps on a genus $g$ Riemann surface with exactly $n$ edges.  Bender and Canfield \cite{BC} showed 
\begin{eqnarray}
M_{n,g} &\sim& t_g n^{5(g-1)/2} 12^n \,\,\, \mbox{as} \,\,\, n \to \infty.
\end{eqnarray}
Many other map classes have similar asymptotics \cite{Gao} with the same sequence of constants $t_g$. In \cite{BC} a recursion is given for these constants the first few of which are:
\begin{eqnarray*}
t_0 = \frac{2}{\sqrt{\pi}} \,\,\,\,\, t_1 = \frac{1}{24} \,\,\,\,\, t_2 = \frac{7}{\sqrt{4320 \pi}}
\end{eqnarray*}
However, there have been some recent improvements in the recursion formula for the $t_g$ \cite{BGR} based on connections to the enumeration of branched coverings of Riemann surfaces \cite{GJ}. More recently \cite{GLM} it has been observed that this new recursion is closely related to the recursion for PI asymptotic coefficients (\ref{PI-rec}). The upshot is that
\begin{eqnarray} \label{t-coeffs}
t_g &=& - \frac{1}{2^{g-2} 6^{\frac{g}{2}}} \cdot \frac{1}{\Gamma(\frac{5g-1}{2})} \alpha_g\\
\nonumber &=& \frac{1}{2^{7g-3}} \cdot \frac{1}{ \Gamma(\frac{5g-1}{2})} a_{3g-1}^{(g)} \,\,\, \mbox{by} \,\,\, (\ref{hop}), 
\end{eqnarray}
which reveals an interesting albeit mysterious link between two prima facie quite different classes of map enumeration problems. 

\appendix{} 
\section{Recurrence Coefficients and their Continuum Limits}
This appendix provides a reasonably self-contained presentation of the relevant results (and outline of their proofs) from 
\cite{EM, EMP}. It also gives the proof of Proposition \ref{d-coeffs}: a new characterization of the coefficients in the continuum Toda equations. 

\subsection{Full Asymptotics of the Recurrence Coefficients}Recall from Section \ref{recurrence} that we have defined $\pi_{n,N}(\lambda,t)$ to be the $n^{th}$ monic orthogonal polynomial with respect to the exponential weight $\exp(-NV(\lambda))$ and that these polynomials satisfy a three-term recurrence relation of the form
\begin{eqnarray}
\pi_{n+1,N}(\lambda) &=& \lambda \pi_{n,N}(\lambda) - b_{n,N}^2(t)  \pi_{n-1,N}(\lambda).
\end{eqnarray}
There is a basic relation between these recurrence coefficients and the tau functions defined in (\ref{GUEEXPNorm}) given by:

\begin{eqnarray}
\label{hirota1} b_{n,1}^2(\theta) &=& \frac{1}{2} \frac{d^2}{d\theta_1^2} \log\left[\tau^2_{n,1}(\theta_1, \theta)\right]_{\theta_1 = 0} \,\, \mbox{where}\\
\label{hirota2} \tau^2_{n,1}(\theta_1, \theta) &=& Z_1^{(n)}(\theta_1,\theta)/Z_1^{(n)}(0,0)\,\, \mbox{and} \\
\label{hirota3} Z_N^{(n)}(t_1, t)  &=& \int \cdots \int \exp\left\{
-N \sum_{j=1}^n \left( \frac{1}{2} \lambda_j^2 + t \lambda_j^{2\nu} +
t_1 \lambda_j \right) \right\} \mathcal{V}(\lambda) d^n \lambda\,,
\\ \nonumber
&&\mbox{where}\; \mathcal{V}(\lambda) = \prod_{j<l} \left| \lambda_j - \lambda_l \right|^2\,.
\end{eqnarray}
The relation (\ref{hirota1}) is called a {\it Hirota formula} \cite{Flaschka} and and stems from the integrable systems theory associated to the Toda Lattice (see Section \ref{continuum}). The connection to orthogonal polynomials comes from the representation of the partition functions (\ref{I.001}) as Hankel determinants of moments for exponential weights \cite{EM, EMP, Szego}.  The definitions on the next two lines are consistent extensions of our earlier definitions. Indeed when $\theta_1, t_1 = 0$, (\ref{hirota2}) and (\ref{hirota3}) are equivalent to (\ref{GUEEXPNorm}) and (\ref{I.001}) respectively.  

The Hirota formula may be extended to more generally scaled recurrence coefficients by making use of the scalings introduced in Section \ref{fs}. 
Indeed, the tau functions are convariant with respect to these scalings so that
\begin{eqnarray}
\tau_{n,1}(\theta_1, \theta) &=& \tau_{n,N}(t_1, t)\, =\, \tau_{n,n}(-s_1, -s) \,\, \mbox{where we define}\\
t_1 &=& 2 \theta_1/ \sqrt{N}  \,\, \mbox{consistent with (\ref{finescale}), and}\\
s_1 &=& -t_1/\sqrt{x}  \,\, \mbox{consistent with (\ref{similarity})}.
\end{eqnarray}
The final transformation to $\tau_{n,n}$ is realized by a variable change $\lambda = \sqrt{x} \hat{\lambda}$ of the eigenvalues in (\ref{hirota3}). 
It then follows from (\ref{hirota1}) that
\begin{eqnarray}
\nonumber \frac{1}{n} b^2_{n,1}(\theta) &=& \frac{1}{2n} \frac{d^2}{d\theta_1^2} \log\left[\tau^2_{n,n}(-s_1,-s)\right]_{s_1 = 0}\\
\nonumber &=& \frac{1}{n^2} \frac{d^2}{ds_1^2} \log\left[\tau^2_{n,n}(-s_1,-s)\right]_{s_1 = 0}; \mbox{moreover,}\\
\label{recjust} \frac{N}{n} b^2_{n,N}(t) = \frac{1}{n} b^2_{n,1}(\frac{-s}{2n^{\nu - 1}}) &=&  \frac{1}{n^2} \frac{d^2}{ds_1^2} \log\left[\tau^2_{n,n}(-s_1,-s)\right]_{s_1 = 0}.
\end{eqnarray}
Theorem \ref{EQMSTHM} may now be applied to the right hand side of (\ref{recjust}) to establish the expansion (\ref{z-exp}).

\subsection{The Continuum Limit of the Toda Lattice Equations for the Recurrence Coefficients} \label{continuum} As the external weight, $V_\nu(\lambda; t)$ changes with $t$, the corresponding orthogonal polynomials, $\pi_{n,N}(\lambda,t)$, also evolve. How they evolve is governed by how the recurrence coefficients, which also depend on $t$, change. It is well-known that the $t$ dependence of the recurrence coefficients is governed \cite{Deift} by a system of nonlinear differential equations known as the {\it Toda Lattice Equations}. There is a different system for each $\nu$. Though different, these different nonlinear systems have a common general form.  The different associated flows on the recurrence coefficients that these different systems induce commute with one another and, in fact, this commutativity is related to the {\it complete integrability} of the Toda Lattices. This feature does not play a direct role in what we present in this paper but it is undoubtedly related to the "universality in $\nu$" that has already been mentioned several contexts.  

In our setting, the Toda Lattice equations take the following form, for each $\nu$:

\begin{eqnarray}\label{beqns}
    \frac{1}{2}\frac{d b^2_n}{d\theta} = \sum_{\{w\}}\left[\prod_{m=1}^{\nu+1}b^2_{n+\ell_m(w)+1} -
    \prod_{m=1}^{\nu+1}b^2_{n+\ell_m(w)}\right]
\end{eqnarray}
where the dependent variable $b^2_n$  stands for $b^2_{n,1}(\theta)$ as defined in (\ref{hirota1}). (A detailed derivation of the form (\ref{beqns}) of the Toda equations from their more standard presentation in the integrable systems literature is given in Section 4.1 of  \cite{EMP}.) The summation in (\ref{beqns}) is taken over the set of walks, denoted $\{ w\}$, of length $2\nu$ on {\bf Z} that start at $+1$ and end at $-1$.  If one visualizes the {\bf Z} lattice along which the walk takes place as a {\it vertical} axis and the {\it steps} of the walk as discrete equally spaced points along a {\it horizontal} axis which serves to order those steps, then the walk is graphed as a zig-zag path comprised of line segments of slope $+45^{\circ}$ (upturn) or $-45^{\circ}$ (downturn). Such a walk is completely determined by specifying where its downturns (of which there are exactly $\nu + 1$) occur. If the discrete variable $\ell$ denotes locations on the vertical axis then, on a given walk, $\ell$ can only range over the interval $[-\nu, \nu]$.  Finally we let $\ell_m(w)$ denote the vertical axis location on the walk $w$ {\it after} its $m^{th}$ downturn; then $\ell_m(w) + 1$ denotes the location {\it before} this downturn. 

In order to study the continuum limit of (\ref{beqns}) we note, from (\ref{recjust}) and (\ref{z-exp}), that these equations contain terms with prima facie different asymptotic gauges:
\begin{eqnarray}
b^2_{n} = b^2_{n}\left(\frac{-s}{2 n^{\nu - 1}}\right) &=& n \left(z_0(s)+\frac{1}{n^2}z_1(s)+\frac{1}{n^4}z_2(s)+\cdots \right)\\
\nonumber b^2_{n+\ell} = b^2_{n+\ell}\left(\frac{-\tilde{s}}{2(n+\ell)^{\nu - 1}}\right) &=& (n+\ell) \left(z_0(\tilde{s})+\frac{1}{(n+\ell)^2}z_1(\tilde{s})+\frac{1}{(n+\ell)^4}z_2(\tilde{s})+\cdots\right)\\
\label{regauge} &=& n \left( w z_0(sw^{\nu -1})+ \dots + \frac{1}{n^{2g}} w^{1-2g} z_g(sw^{\nu -1})  + \cdots \right) ,
\end{eqnarray}  
where $w = 1 + \frac{\ell}{n}$ as defined in (\ref{gaugevar}). The introduction of the latttice scaling variable $w$ in (\ref{regauge}) allows us to analyze all the terms appearing in (\ref{beqns}) with respect to the same asymptotic gauge. Since $\ell \in [-\nu, \nu]$ and $\nu$ is fixed,
one may assume that $\ell << n$. Then one can study the limit as $n \to \infty$ of the equations (\ref{beqns}), with the substitutions (\ref{regauge}), as one would study the continuum limit of a numerical scheme or a molecular chain with analytic potential. To carry this out, we introduce a more compact notation for (\ref{regauge}) as
\begin{eqnarray}
\nonumber f(s,w) &=& f_0(s,w) + \frac{1}{n^2}f_1(s,w)+ \cdots + \frac{1}{n^{2g}}f_g(s,w) + \cdots,\,\,\, \textrm{where}\\
\nonumber  f_g(s,w) &=&  w^{1-2g}z_g(w^{\nu-1}s) \,\,\, \textrm{so that}\\
 \label{scheme}  b^2_{n+\ell} &=& n f\left(s,1 + \frac{\ell}{n}\right) \,\,\, \textrm{and in particular}\\
\nonumber  n^{-1} b^2_{n} &=&  f(s,1) = z_0(s)+\frac{1}{n^2}z_1(s)+\frac{1}{n^4}z_2(s)+\cdots\,\,\, \textrm{and}\\
\label{dscheme} \frac{1}{2}\frac{d b^2_n}{d\theta} &=& 2 n^{\nu-1} \frac{1}{2}\frac{d b^2_n}{ds}  = \left. n^\nu \frac{d}{ds}f(s,w) \right|_{w=1} 
\end{eqnarray}
Finally, substituting (\ref{scheme}) and (\ref{dscheme}) into (\ref{beqns}) and Taylor expanding in $\frac{\ell_m}{n}$ (and $\frac{\ell_m + 1}{n})$ one arrives at the basic continuum equations
\begin{eqnarray}
\nonumber \left. \frac{d}{ds}f(s,w)\right|_{w=1} &=& \sum_{\{w\}} nf^{\nu+1}\left\{\prod_{m=1}^{\nu+1}\left[ 1 + \frac{f_w}{f}\left(\frac{\ell_m +1}{n}\right) + \frac{f_{w^{(2)}}}{2f}\left(\frac{\ell_m +1}{n}\right)^2 + \cdots + \frac{f_{w^{(h)}}}{h! f}\left(\frac{\ell_m +1}{n}\right)^h + \cdots \right]\right. \\
 && \hspace{0.59in} - \left. \prod_{m=1}^{\nu+1}\left[ 1 + \frac{f_w}{f}\frac{\ell_m}{n} + \frac{f_{w^{(2)}}}{2f}\left(\frac{\ell_m}{n}\right)^2 + \cdots + \frac{f_{w^{(h)}}}{h! f}\left(\frac{\ell_m}{n}\right)^h + \cdots
   \right]\right\} \label{HCAAA03}
\end{eqnarray}
from which Theorems \ref{cont} and \ref{emp} can be deduced. 

\subsection{Coefficient Formulae for the Continuum Toda Equations} \label{forcing-coeffs}

We now provide a proof of Proposition \ref{d-coeffs}. Recall from Theorem \ref{cont} that, using multi-index notation, 
\begin{eqnarray*}
F_g^{(\nu)} &=&  \sum_{
\begin{array}{c}
|\lambda|   = 2g+1   \\
\ell(\lambda)   \leq  \nu + 1    
\end{array}
} \frac{d_\lambda^{(\nu,g)}}{\prod_j r_j(\lambda)! } f^{\nu -\ell(\lambda)+1} \frac{1}{\lambda!} \frac{\partial^{|\lambda|}f}{\partial w^\lambda}
\end{eqnarray*}
These coefficients have the character of correlation functions for certain tied random walks. A description based on this perspective is presented in \cite{EMP}. In the following derivation, that description will be used but not explained in full detail. 
\medskip

 {\bf Proof of Proposition \ref{d-coeffs}.}  The coefficients $d_\lambda^{(\nu,g)}$ are directly related to the expression of the Toda Lattice equations for the $b_k^2$, in terms of tied walks on a one dimensional lattice as described in \cite{EMP}. The walks in question can be reduced to considering walks of length $2\nu$ on {\bf Z} that start at $+1$ and end at $-1$.  If one visualizes the {\bf Z} lattice along which the walk takes place as a {\it vertical} axis and the time steps as discrete equally spaced points along a {\it horizontal} axis, then the walk is graphed as a zig-zag path comprised of line segments of slope $+45^{\circ}$ (upturn) or $-45^{\circ}$ (downturn). We label the locations on the vertical axis by $\ell$ and those on the horizontal axis by $i$. For the walks under consideration, $\ell$ ranges over the interval 
$[-\nu, \nu]$ and $i$ ranges from  $1$ to $2\nu$. Such a walk is completely determined by specifying when its downturns (of which there are exactly $\nu + 1$) occur. We will let $i_j$, for $j \in \{1,\dots \nu+1\}$, denote the step at which the $j^{th}$ downturn takes place. We also let $\ell_j$ denote the location of the path {\it after} the  $j^{th}$ downturn (then $\ell_j+1$ denotes the location of the path {\it before} the  $j^{th}$ downturn). 

Given this set-up, and based on the derivation given in \cite{EMP}, the coefficients can be expressed as
\begin{eqnarray}
\nonumber d_\lambda^{(\nu,g)} &=&  \sum_{\mbox{walks}} \left[m_\lambda\left(\ell_1 +1, \dots, \ell_{\nu+1}+1\right) - m_\lambda\left(\ell_1, \dots, \ell_{\nu+1}\right)\right] \\
&\label{d-formula}&\\
\nonumber &=& \sum_{ i_1 < \dots < i_{\nu+1}} \left[m_\lambda\left(i_1, i_2-2, \dots, i_{\nu+1} -2\nu \right) - m_\lambda\left(i_1-1, i_2-3, \dots, i_{\nu+1}-2\nu-1\right)\right]. 
\end{eqnarray}
We now re-express the sequence of downturn steps as a restricted partition as follows
\begin{eqnarray*}
\mu &=& (\mu_1, \dots, \mu_j, \dots, \mu_{\nu+1}) = (i_{\nu+1}, \dots, i_{\nu+1 -(j -1)}, \dots, i_1).
\end{eqnarray*}
We also define the partition 
\begin{eqnarray*}
\eta &=& (2\nu, 2\nu-2, \dots, 2,0)
\end{eqnarray*}
and for notational convenience we define the difference
\begin{eqnarray*}
\mu - \eta &=& (i_{\nu+1}-2\nu, i_\nu -2\nu +2, \dots, i_2 - 2, i_1).
\end{eqnarray*}
Since $m_\lambda$ is a symmetric polynomial we may rewrite (\ref{d-formula}) as 
\begin{eqnarray*}
d_\lambda^{(\nu,g)} &=&  \sum_{\mu \in \mathcal{R}_{(\nu+1, \nu, \dots, 2,1)}^{(2\nu, 2\nu-1, \dots, \nu)}} \left[m_\lambda(\mu-\eta) - m_\lambda(\mu-\eta - (1,\dots, 1))\right] 
\end{eqnarray*}
where 
$$
\mathcal{R}_{(\nu+1, \nu, \dots, 2,1)}^{(2\nu, 2\nu-1, \dots, \nu)} = \left\{ \mu \in \mathcal{R}: (\nu+1, \nu, \dots, 2,1) \subseteq \mu \subseteq (2\nu, 2\nu-1, \dots, \nu)\right\}.
$$
Note further that
\begin{eqnarray*}
\left(-(\nu-1), -\nu, \dots, 0,1\right) \subseteq &\mu - \eta& \subseteq  \left( 0, 1, \dots, \nu-1, \nu \right)\\
\left(-\nu, -\nu+1, \dots, -1,0\right) \subseteq &\mu - \eta - (1, 1, \dots, 1)& \subseteq  \left(-1, 0, \dots, \nu-2, \nu-1\right).
\end{eqnarray*}
It follows that $\mu - \eta - (1, 1, \dots, 1)$ ranges over the same set of sequences as $\mu - \eta$, but with all signs reversed. Hence, the sum in the last expression for $d_\lambda^{(\nu,g)}$ may be rearranged to read
\begin{eqnarray*}
d_\lambda^{(\nu,g)} &=& \sum_{\mu \in \mathcal{R}_{(\nu+1, \nu, \dots, 2,1)}^{(2\nu, 2\nu-1, \dots, \nu)}} \left[m_\lambda\left(\mu-\eta\right) - m_\lambda\left(-(\mu-\eta)\right)\right] \\
&=&  \sum_{\mu \in \mathcal{R}_{(\nu+1, \nu, \dots, 2,1)}^{(2\nu, 2\nu-1, \dots, \nu)}} \left[m_\lambda\left(\mu-\eta\right) - (-1)^{|\lambda|}m_\lambda\left(\mu-\eta\right)\right] \\
&=& \sum_{\mu \in \mathcal{R}_{(\nu+1, \nu, \dots, 2,1)}^{(2\nu, 2\nu-1, \dots, \nu)}} 2 \,\, m_\lambda\left(\mu-\eta\right),
\end{eqnarray*}
where the last line follows because $|\lambda| = 2g+1$ which is odd. $\Box$
\begin{remark}
The formula for $d_\lambda^{(\nu,g)}$ in Proposition \ref{d-coeffs} differs significantly from that given in \cite{EMP}. Both versions have their respective merits. The one introduced here is, for example, useful for growth estimates of these coefficients with respect to their parameters. 
\end{remark}

\section{Large-Time Behavior  of the One-Point Correlation Function} \label{largetime} In this section we are going to discuss the large $N$ behavior of the one-point correlation function for eigenvalues of $N \times N$ random hermetian matrices in the vicinity of exponential weights $V_t(\lambda) \doteq V_\nu(\lambda; t)$ for which $t$ is very large.  We refer to this as {\it large-time} behavior because $t$ has the interpretation of a dynamic variable of the $2 \nu^{th}$ Toda flow.  This discussion applies to the determination of the large-time behavior of the coefficients $z_g$ and $e_g$.  The analysis of the large $N$ behavior of the one-point correlation function that was carried out in \cite{EM} was done explicitly only for values of $t$ close to zero. 
However, as was pointed out in that paper, the range of validity of the large $N$ expansion derived there was by no means restricted to a small neighborhood of $t=0$.   Indeed, the expansion can be constructed with uniform validity in an open complex neighborhood of any $t > 0$ as long as the equilibrium measure $\mu_{V_t}$ for the weight $V_t$ at that value of $t$, satisfies the following conditions: 
\begin{enumerate}
\item[(a)] The  support of  $\mu_{V_t}$ is a single  interval of the form $[-\beta,\beta]$;  
\item[(b)] $\mu_{V_t}$ satisfies the variational equations of (\ref{VProb}); 
\item[(c)]  $\mu_{V_t}$ vanishes like a square root at both endpoints of its support. 
\end{enumerate} 
In Section \ref{uniform} we will show that these conditions are indeed satisfied for all $t > 0$ by explicitly writing down the equilibrium measure for all these values of $t$. It then follows, from analytic continuation and uniqueness of power series expansions for analytic functions, that the coefficients of the large $N$ expansions of logarithms of tau functions (\ref{GUEEXPNorm}) and their derivatives, at these values of $t$, coincide with those constructed near $t=0$ (such as $e_g$ and $z_g$). 

However, for our applications we require more. We need to know the limiting values of these coefficients as $t \to \infty$. To accomplish this one should actually build the asymptotic expansions of correlation functions in a neighborhood of $t = \infty$. In section \ref{RHlarge} we indicate how the constructions of \cite{EM} may be adapted to this large-time regime.

\subsection{The Equilibrium Measure for $t  > 0$} \label{uniform}
We present here an explicit expression for the equilibrium measure (\ref{eqmeas}) that is uniformly valid for all $t \geq 0$.  To accomplish this, we rescale the domain variable  $\lambda$ so that the support of the measure remains fixed as $t$ varies:
\begin{eqnarray}
\label{z0-uniform} \lambda &=& 2 \sqrt{z_0} \eta\,\, \mbox{where}\\
\nonumber z_0 &=& z_0(-t)\,\, \mbox{as defined in (\ref{z0})}.
\end{eqnarray}
 
\begin{prop} The equilibrium measure for $t \geq 0$ is given, under the variable dilation (\ref{z0-uniform}), by the following expression which is a convex linear combination, for $z_0 \in [0, 1]$, of two probability measures on $[-1,1]$:  
\begin{eqnarray} 
\nonumber d\mu_{V_t}(\eta) &=& \frac{2}{\pi} \chi_{(-1, 1)} (\eta) \left\{ z_0 
+ (1- z_0)\left[ \frac{(2\eta)^{2\nu-2}}
{\begin{pmatrix}
  2 \nu - 1\\
   \nu - 1\\
\end{pmatrix} } \right. \right. \\
\label{uniform-meas} &+& 
\left. \left. \sum_{j=1}^{\nu - 1}  
2 \frac{ \begin{pmatrix}
 2j - 1   \\
   j - 1
\end{pmatrix}}
{\begin{pmatrix}
  2 \nu - 1\\
   \nu - 1\\
\end{pmatrix} } 
(2\eta)^{2\nu -2 -2j}\right] \right\} \sqrt{(\eta+1)(1-\eta)} \,\, d\eta.
\end{eqnarray}
This expression depends on $t$ {\it only} through the parameter $z_0$. When $z_0 = 1$ this corresponds to the weight for $t=0$ and the measure has the density of the Wigner semicircle law; when $z_0 = 0$ this corresponds to $t=\infty$ which we think of as the limiting asymptotic measure for large $t$. This limiting density also gives a probability measure supported on $[-1,1]$. It is in fact the density of the equilibrium measure for the pure monomial weight
\begin{eqnarray} \label{monpot}
V_\infty(\eta) &=& \frac{2^{2\nu}}{c_\nu} \eta^{2\nu}. 
\end{eqnarray}
\end{prop}

{\bf Proof.} Pinning down an explicit expression for (\ref{eqmeas}) amounts to determining the coefficients of the polynomial $h(\lambda)$. We already know $\beta$, from (\ref{beta}), explicitly in terms of $z_0$. The essential ingredients for calculating 

\begin{equation} \label{hpoly}
h(\lambda) =  1 + \sum_{j=0}^{\nu - 1} h_j\lambda^{2j} 
\end{equation}
were developed in section 3 of \cite{EMP}.  We recall those elements here.

Define the sequence $\{ v_j\}_{j=0}^\infty$ by
\begin{equation} \nonumber \index{$v_j$}
\frac{ \sqrt{\lambda^2 - \beta^2}}{\lambda} = 1 -
\sum_{i=0}^{\infty}
  v_i \frac{1}{\lambda^{2i+2}};
\end{equation}
whose Taylor coefficients can be computed to be
\begin{equation} \label{v_k}
v_i = \frac{1}{4^i} \binom{2i-1}{i-1} \frac{\beta^{2i+2}}{i+1},
\end{equation}
for $i > 0$ with $v_0$ defined to be $\beta^2/2$. The coefficients of $h$ are directly expressed in terms of the $\{ v_j\}$ as 

\begin{eqnarray} \label{h_j}
h_j &=& 4 \nu (\nu - j) t \frac{ v_{\nu -1 -j} }{\beta^2 }\\
\nonumber &=& 4 \nu (\nu - j) \frac{1-z_0}{c_\nu z_0^\nu} \frac{ v_{\nu -1 -j} }{\beta^2 }\\
\nonumber &=& 2
\frac{ \begin{pmatrix}
 2 \nu - 2j - 3   \\
   \nu - j - 2
\end{pmatrix}}
{\begin{pmatrix}
  2 \nu - 1\\
   \nu - 1\\
\end{pmatrix} } 
\frac{(1-z_0)}{z_0^{j+1}},\,\, \mbox{for}\,\, j < \nu - 1;\\
\nonumber h_{\nu - 1} &=&  \frac{ 1}
{\begin{pmatrix}
  2 \nu - 1\\
   \nu - 1\\
\end{pmatrix} }
\frac{(1-z_0)}{z_0^{\nu}}.
\end{eqnarray}
Equation (\ref{II.005}) was used to eliminate $t (= -s)$ and $c_\nu$ was replaced by its definition from (\ref{z0}). The exact expression (\ref{uniform-meas}) now follows by substituting (\ref{hpoly}) with the coefficients (\ref{h_j}) into (\ref{eqmeas}), using (\ref{beta}) to re-express $\beta$ in terms of $z_0$, changing variables according to (\ref{z0-uniform}) and, finally, changing the index of summation from $j$ to $\nu - j$.

It is an exercise to check that $\mu_{V_\infty}$ is a probability measure by directly integrating the density (\ref{uniform-meas}), for $z_0 = 0$, over $[-1,1]$. (It is manifest that this density is positive on $(-1,1)$.)  The equilibrium measure for monomial weights such as (\ref{monpot}) is explicitly calculated on page 183 of \cite{Deift}. One may check, by direct comparison, that (\ref{uniform-meas}), with $z_0 = 0$, is indeed the equilibrium measure for (\ref{monpot}). The next section will give another perspective on why this is the case.  
$\Box$

\subsection{The Riemann-Hilbert Problem at Infinity} \label{RHlarge} 

The prior results (Theorems \ref{EQMSTHM}, \ref{II.002}, \ref{cont}) cited in Sections \ref{intro} and \ref{background} , and on which the results of this paper fundamentally depend,  all derive from a detailed asymptotic analysis of the spectral density for eigenvalues of random Hermetian matrices that was carried out in \cite{EM}.  Explicitly this spectral measure is defined as the expectation
\begin{eqnarray}
\rho_1^{(N)}(t,\lambda) &=& \frac{d}{d\lambda} \mathbb{E}_{\mu_t} \left(\frac{1}{N}\# \left\{j: \lambda_j \in (-\infty, \lambda)\right\}\right)\,\, \mbox{w.r.t. the probability density}\\
\nonumber d\mu_t &=&\frac{1}{Z_N(t)} 
\exp{ \left\{-N^{2}\left[\frac{1}{N} \sum_{j=1}^{N} V_t(\lambda_{j})  - \frac{1}{N^{2}} \sum_{j\neq \ell} \log{|\lambda_{j} -
\lambda_{\ell} | } \right] \right\} } d^{N}\lambda. 
\end{eqnarray}
This has a remarkable expression as a {\it one-point correlation function} in terms of the Wronskian determinant
\begin{eqnarray} \label{3.003}
\rho_{1}^{(N)}(\lambda) =  K_N(\lambda, \lambda) = \frac{e ^{-N V_t(\lambda)}}{-2 \pi i N} \left[
Y_{11}'(\lambda) Y_{21}(\lambda) - Y_{11}(\lambda) Y_{21}'(\lambda) \right]
\end{eqnarray}
of the following Riemann-Hilbert (RH) problem  for the $2 \times 2$ matrix $Y(\lambda)$:

\begin{itemize}
\item $Y$ analytic in $\mathbb{C}\backslash\mathbb{R}$,
\item $Y = \bigl( I + \, \mathcal{O} (\frac{1}{\lambda})\bigr)\, \lambda^{N\sigma_3}$,
\item $Y$ has H\"older continuous boundary values $Y_\pm$, from above and below respectively, along $\lambda \in \mathbb{R}$,
\item $Y_{+} = Y_{-} \Bigl( \begin{smallmatrix} 1 & e^{-NV_t(\lambda)}\\ 0 & 1
\end{smallmatrix}\Bigr)$, for $\lambda \in \mathbb{R}$,
\end{itemize}
where $\sigma_3$ is the diagonal Pauli matrix referred to by this standard notation.  This problem is directly related to a RH problem for orthonormal polynomials with weight $V_t$. We will not need to go into that here, but refer the reader to \cite{EM} for more details and references.

The relevance of the one-point function to the problems discussed in this paper can be seen from the fact that
\begin{eqnarray}
Z_N({ t}\, ) = Z_N({ 0}) \exp{ \left\{ -N^{2}\int_0^{ t} \int_{\mathbb{R}}
\rho_1^{(N)} (t,\lambda ) \, \lambda^{2\nu} {d\lambda} dt \right\}}.
\end{eqnarray}
This falls within the more general class of problems which study the large $N$ limits of integrals of the form
\begin{eqnarray} \label{general}
\int_{\mathbb{R}} \rho_1^{(N)} (t,\lambda ) \, f(\lambda) {d\lambda}
\end{eqnarray}
where $f(\lambda)$ is a general $C^\infty$ with no worse than polynomial growth at infinity. 

We now want to study how the RH problem for $Y$ behaves in the limit as $t \to \infty$. Under the change of variables (\ref{z0-uniform}) the problem for $Y$ transforms to an equivalent problem 
\begin{itemize}
\item $Z$ analytic in $\mathbb{C}\backslash\mathbb{R}$,
\item $Z = \bigl( I + \, \mathcal{O} (\frac{1}{\eta})\bigr)\, \eta^{N\sigma_3}$,
\item $Z$ has H\"older continuous boundary values $Z_\pm$ along $\eta \in \mathbb{R}$,
\item $Z_{+} = Z_{-} \Bigl( \begin{smallmatrix} 1 & e^{-N\tilde{V}_t(\eta)}\\ 0 & 1
\end{smallmatrix}\Bigr)$, for $\lambda \in \mathbb{R}$,
\end{itemize}
where  
\begin{eqnarray}
\nonumber \tilde{V}_t(\eta) &=& 2 z_0 \eta^2 + 2^{2\nu} t z_0^\nu \eta^{2\nu}\\
\nonumber &=& 2 z_0 \eta^2 + 2^{2\nu} \frac{1-z_0}{c_\nu z_0^\nu} z_0^\nu \eta^{2\nu}\\
                     &=&  z_0\,\,\, 2\eta^2 + (1-z_0)\,\,\, \frac{2^{2\nu}}{c_\nu}  \eta^{2\nu}.
\end{eqnarray}
Equation (\ref{II.005}) was used to eliminate $t (=-s)$.  We see that as $t \to \infty, z_0 \to 0$ and the weight limits to $\frac{2^{2\nu}}{c_\nu}  \eta^{2\nu}$, consistent with (\ref{monpot}). This establishes that the equilibrium measures  (\ref{eqmeas}), for the variational problems (\ref{VProb}), converge as $t \to \infty$  to the equilibrium measure for the variational problem with monomial weight (\ref{monpot}) that is associated to the RH problem for $Z$ at $z_0 = 0$. Given this, all of the analysis of \cite{EM} to derive the full asymptotic expansion for   
$\rho_1^{(N)}$ uniformly valid for positive $t$ near $0$, carries over {\it mutatis mutandis} to give the asymptotic expansion of this one-point function near $t = \infty$:
\begin{thm} \label{infexpB}
There is a $z_0^* \in [0,1]$ such that for all $z_0 \in [0, z_0^*)$, the following asymptotic expansion is valid uniformly in $z_0$:
\begin{eqnarray*}
\int_{-\infty}^{\infty} f(\lambda) \rho_1^{(N)} (t(z_0),\lambda) d\lambda &=& f_{0} + N^{-2} f_{1} + N^{-4} f_{2} + \cdots
\end{eqnarray*}
provided that the function $f(\lambda)$ is $C^\infty$ and grows no faster than a polynomial as $|\lambda| \to \infty$. The coefficients depend analytically on $t$ for $z_0(-t) \in [0, z_0^*)$, and the asymptotic expansion may be differentiated term by term.
\end{thm}

This construction is reminiscent of the small amplitude limit for action-angle variables in Hamiltonian mechanics. It is, in effect, a one-point compactification at infinity of the family of variational problems (\ref{VProb}), their respective unique solutions, and associated RH problems all parametrized by $t \in [0, \infty)$. This parametrization is now naturally referenced to the compact interval $0 \leq z_0 \leq 1$.  
\medskip

Finally we apply this result to extend the domain of uniform validity in $t$ for Theorems  \ref{EQMSTHM} and \ref{II.002}. 
\begin{thm} \label{extension}
There is a constant $\Delta >0$ such that for (complex) $t$ with $\Re(t) \geq 0$, $|\Im(t)| < \Delta$ one has a uniformly valid asymptotic expansion
\begin{eqnarray}
\label{I.002B} \ \ \ \log \tau^2_{N,N}(t) =
N^{2} e_{0}({t}) + e_{1}({{t}}) + \frac{1}{N^{2}}e_{2}(t) + \cdots
\end{eqnarray}
as $N \to \infty$. Also,
the recurrence coefficients for the monic orthogonal polynomials with weight $\exp(-N V(\lambda))$ have a full asymptotic expansion, uniformly valid for (complex) $t$ with $\Re(t) \geq 0$, $|\Im(t)| < \Delta$ , of the form
\begin{eqnarray}\label{z-expB}
b_{N,N}^2(t)  &=&  z_0(-t)+\frac{1}{N^2}z_1(-t)+\frac{1}{N^4}z_2(-t)+\cdots 
\end{eqnarray} 
as $N \to \infty$.  The meaning of these expansions is:  if you keep terms up to order
$N^{-2h}$, the error term is bounded by $C N^{-2h-2}$, where the constant $C$ is independent of $t$ in the domain \\ 
$\left\{(\Re t \geq 0;   -\Delta < \Im t < \Delta\right\}$.  Moreover, in this domain, for each $\ell$, the functions $e_{\ell}(t)$ and  $z_{\ell}(-t)$ are analytic functions of $t$ and the asymptotic expansion of derivatives of $\log{ \left( Z_{N}(t) \right)}$ and $b_{N,N}^2(t)$ may be calculated via 
term-by-term differentiation of the series.
\end{thm}

{\bf Proof.} For any $t \in [0, \infty)$ the methods of \cite{EM} enable one to construct asymptotic expansions of the form (\ref{I.002B}) and (\ref{z-expB}) which are uniformly valid in a complex neighborhood of that $t$ intersected with the half plane $\Re t \geq 0$. By Theorem \ref{infexpB} we also have such a neighborhood around $t=\infty$. Consider an open covering of $[0,\infty]$ by such complex neighborhoods which corresponds to an open cover of $z_0 \in [0,1]$. By compactness of this latter interval there is a finite sub-cover which in turn corresponds to a finite open cover of  $t \in [0,\infty)$ for which the genus expansion (\ref{I.002}) and the recurrence coefficient expansion (\ref{z-exp}) is uniformly valid. Since the cover is finite, a width $\Delta$ in Theorem \ref{extension} can be determined. $\Box$

The uniform validity of these expansions in a semi-infinite strip of the $t$-plane is important for the results of this paper. See, in particular, section \ref{z0-reg} and the end of section \ref{partfcn-asymp}.

\bibliographystyle{amsplain}

\end{document}